\documentclass[12pt,reqno]{amsart}

\usepackage[OT1]{fontenc}
\usepackage{amsthm,amsmath,amsfonts,amssymb,amsaddr}
\usepackage{xr-hyper}
\usepackage{xr}
\usepackage[colorlinks=true,linkcolor=black, citecolor=blue, urlcolor=blue]{hyperref}
\usepackage{mathrsfs}
\usepackage{ascmac}
\usepackage{bm}
\usepackage{fancybox}
\usepackage{graphicx}
\usepackage{xcolor}
\usepackage{tabularx}
\usepackage{subcaption}
\usepackage{algorithm2e}

\usepackage{fancyhdr}
\usepackage[semicolon, sort,comma,authoryear,round]{natbib}
\usepackage{fullpage}
\usepackage{newtxtext,newtxmath}
\usepackage{csvsimple}

\newtheorem{theorem}{Theorem}

\theoremstyle{definition}

\newcommand{\R}{\mathbb{R}}

\newcommand{\Ep}{\mathbb{E}}
\renewcommand{\Pr}{\mathbb{P}}

\newcommand{\T}{\mathrm{\scriptscriptstyle T}}

\renewcommand{\hat}{\widehat}

\newcommand{\given}{\,|\,}

\DeclareMathOperator{\Var}{Var}

\makeatletter
\newcommand*{\addFileDependency}[1]{
\typeout{(#1)}
\@addtofilelist{#1}
\IfFileExists{#1}{}{\typeout{No file #1.}}
}\makeatother

\newcommand*{\myexternaldocument}[1]{%
\externaldocument{#1}%
\addFileDependency{#1.tex}%
\addFileDependency{#1.aux}%
}
\myexternaldocument{supplementary}

\title{Assessing Spatial Disparities: A Bayesian Linear Regression Approach}
\author{Kyle Lin Wu and Sudipto Banerjee}
\date{\today}

\begin{document}

\begin{abstract}
Epidemiological investigations of regionally aggregated spatial data often involve detecting spatial health disparities among neighboring regions on a map of disease mortality or incidence rates. Analyzing such data introduces spatial dependence among health outcomes and seeks to report statistically significant spatial disparities by delineating boundaries that separate neighboring regions with disparate health outcomes. However, there are statistical challenges to appropriately define what constitutes a spatial disparity and to construct robust probabilistic inferences for spatial disparities. We enrich the familiar Bayesian linear regression framework to introduce spatial autoregression and offer model-based detection of spatial disparities. We derive exploitable analytical tractability that considerably accelerates computation. Simulation experiments conducted on a county map of the entire United States demonstrate the effectiveness of our method, and we apply our method to a data set from the Institute of Health Metrics and Evaluation (IHME) on age-standardized US county-level estimates of lung cancer mortality rates.
\end{abstract}

\maketitle

\section{Introduction}\label{sec: intro}
We develop statistical methods for detecting local spatial disparities on maps. Disparities in health outcomes are a rapidly evolving field within public health \citep[][]{rao2023} and spatial variation in health outcomes affects how disparities manifest on disease maps. Spatial data analysis in epidemiological investigations is extensively documented \citep[][]{waller2004applied, waller2010handbook, lawson2013statistical, lawson2016handbook}. Inference on disparities proceeds from ``difference boundaries'' that represent significant differences between neighbors and delineate regions with significantly different disease mortality and incidence rates, helping to improve the geographic allocation of health care resources and guide future investigation into local determinants. Boundary detection enables precise identification of health event clusters and sudden changes in latent or observed environmental exposures, which is crucial for defining sampling protocols across a diverse range of health environments and planning targeted interventions \citep{copelandRolePublicHealth2010, jacquezGeographicBoundaryAnalysis2010}. For example, \cite{barboza-salernoPushingBoundaryChild2025} apply boundary detection to identify local socioeconomic inequalities and estimate their association with neighborhood child mortality.
In geographic analysis, the problem is often called ``wombling'' (see \citeauthor{jacquez2003a}, \citeyear{jacquez2003a}a, \citeyear{jacquez2003b}b; \citet[][]{lu2005bayesian, li2011mining, ma2007bayesian, fitzpatrick2010} for some algorithmic approaches with applications). 
Probabilistic model-based approaches include developments in \cite{lu2007bayesian, ma2010hierarchical, li2015bayesian, banerjee2012bayesian, hanson2015spatial, leeBoundaryDetectionDisease2012, corpas-burgos2020serra, gaoSpatialDifferenceBoundary2023} and \cite{aielloDetectingSpatialHealth2025}. 

The aforementioned articles present an impressive range of methods, but are broadly classified into two groups. The first embraces hierarchical models and Bayesian inference for the spatial adjacency matrix by modeling its elements. The second endows the spatial random effects with a probability law that accommodates nonzero probabilities on neighboring spatial random effects being equal (often using an areal Dirichlet process or some adaptation thereof), which are computed as posterior probabilities. Both groups introduce spatial dependence and aim to balance the underlying spatial similarities with the ability to detect differences among neighbors. Our current contribution offers a computationally scalable and analytically tractable approach to this problem. Instead of modeling spatial effects as realizations of a Dirichlet process, which struggles to learn from the data and impedes convergence of iterative estimation algorithms, we devise fully model-based Bayesian inference within an augmented spatial linear regression framework. More specifically, we treat spatial disparities as a multiple comparison problem within a Bayesian inferential paradigm by testing for differences in spatial random effects between neighboring regions. Subsequently, we use posterior probabilities of such differences to detect spatial disparities while controlling a Bayesian false discovery rate or FDR \citep[][]{mullerFDRBayesianMultiple2006}. Controlling the false discovery rate is preferred to controlling the family wise error rate or false positive rate because it offers increased power and flexibility \citep{catelanMultipleTestingDisease2010, glickmanFalseDiscoveryRate2014} and has been previously applied in analyses of health disparities \citep{tianIdentificationRacialDisparities2010, slutskeExplainingCOVID19Related2023}. Further developments of Bayesian FDR for multiple testing in spatial settings include \cite{ventrucciMultipleTestingStandardized2011, sunFalseDiscoveryControl2015, tanseyFalseDiscoveryRate2018}.

Benefits of our approach include clearer interpretation of spatial effects, computational simplicity, accounting for small differences in the values of neighboring spatial effects in determining spatial disparity, and providing improved sensitivity and specificity for detecting boundary differences \citep[as seen in][]{gaoSpatialDifferenceBoundary2023}. Our parametric approach is suitable for applications where one disease map is of interest and lack of replication hinders non-parametric methods \citep[such as in][]{slackGeographicConcentrationUs2014, jagai2017county, rodriguez2018county}.    

We exploit closed-form distribution theory to account for the differences between the values of the spatial effects and evaluate posterior probabilities of such differences exceeding a certain threshold. We extend \cite{lu2005bayesian} and \cite{fitzpatrick2010}, who use only posterior means or spatial residuals to detect differences, to fully model-based inference. We provide an outline for the rest of the article. Section~\ref{sec: lin_reg} discusses the problem of detecting differences between fixed effects in a linear model before we extend to spatial random effects. Section~\ref{sec: new_methods} develops the notion of a difference boundary, demonstrates a meaningful connection to traditional hypothesis testing in the simple linear regression model, and considers extensions to a generalized linear model. Section~\ref{sec:simulation_experiments} evaluates our methods alongside an existing Bayesian nonparametric approach using simulated data over a map of Californian counties followed by an analysis of 
lung cancer mortality data in Section~\ref{sec:application}. Section~\ref{sec: discussion} concludes with a brief discussion. 

\section{Assessing Contrasts in Bayesian Linear Regression}\label{sec: lin_reg}

\subsection{Random Effects under an Augmented Linear System}\label{subsec:ME_model}

We follow the adaptation by \cite{rieblerIntuitiveBayesianSpatial2016} of the ``BYM'' model \citep{besagBayesianImageRestoration1991} to ensure identifiable parameters and provide better interpretation. The ``BYM2'' model is
\begin{gather}\label{eqn:BYM2_model}
    y = X\beta + \gamma + \eta, \;\;
    \gamma = \sigma\sqrt{\rho}\phi, \;\; \phi \sim \mbox{N}_n(0, V_\phi), \;\; \eta = \sigma\sqrt{1 - \rho}v, \;\; v \sim \mbox{N}_n(0, I_n) 
\end{gather}
where $\phi$ is an $n \times 1$ vector of random effects, 
$V_\phi$ is a known positive definite matrix such that $\pi(\phi)$ is proper. The parameter $\sigma^2$ captures the overall error variance if $c$ is set so that the geometric mean of the marginal prior variances is equal to one, and $\rho$ represents the proportion of total variance attributed to the spatial structure \citep{rieblerIntuitiveBayesianSpatial2016}. For subsequent analysis, we specify a conditionally autoregressive (CAR) structure, one of the most popular spatial models used in areal data analysis. Here, $\pi(\gamma_{i} \given \sigma^2, \gamma_{j}, j \neq i) = \mbox{N}(\alpha \sum_{j}w_{ij} \gamma_j / w_{i+}, c\sigma^2/w_{i+})$, where $w_{ij}$ is the binary adjacency indicator for regions $i$ and $j$, $\alpha$ is a spatial smoothing parameter, $c$ is a fixed scaling constant and $w_{i+} = \sum_{j} w_{ij}$ is the number of neighbors of region $i$. The conditional distribution of each spatial residual is based on a linear combination of its neighbors. By Brook's lemma \citep{besagBayesianImageRestoration1991}, this equates to setting $V^{-1}_\phi = c(D_W - \alpha W)$ in \eqref{eqn:BYM2_model}, where $W$ is the adjacency matrix, $D_W$ is diagonal with the diagonal elements $D_{W, ii} = w_{i+}$ \citep[][Chap. 4]{banerjeeHierarchicalModelingAnalysis2015}. The density $\pi(\phi)$ is proper when $\alpha \in (1 / \lambda_{min}, 1)$, where $\lambda_{min}$ is the smallest eigenvalue of $D_W^{-1/2}WD_W^{-1/2}$. Since $\rho$ already controls the amount of spatial smoothing in \eqref{eqn:BYM2_model}, we set $\alpha = 0.99$ in the following sections to improve the identifiability of spatial effects $\phi$, which we consider as potential indicators of latent factors driving local health inequity.  

If $\pi(\gamma, \beta, \sigma^2, \rho) =  \mbox{N}(\gamma \given 0, \sigma^2\rho V_\phi) \times \mbox{N}(\beta \given M_0 m_0, \sigma^2 M_0) \times \text{IG}(\sigma^2 \given a_{\sigma^2}, b_{\sigma^2}) \times \pi(\rho)$ is the prior for \eqref{eqn:BYM2_model}, then the joint distribution arises from the augmented linear model
,
\begin{equation}\label{eqn:augmented_system}
    \underbrace{\begin{bmatrix}
        y \\
        M_0m_0 \\
        0_{n} 
    \end{bmatrix}}_{y_\star} =
    \underbrace{\begin{bmatrix}
        X & I_n \\
        I_p & 0_{p\times n} \\
        0_{n \times p} & I_n
    \end{bmatrix}}_{X_\star}
    \underbrace{\begin{bmatrix}
        \beta \\
        \gamma
    \end{bmatrix}}_{\gamma_\star}
    + \underbrace{\begin{bmatrix}
        \eta \\
        \eta_{\beta} \\
        \eta_{\phi}
    \end{bmatrix}}_{\eta_{\star}}, \;\;
    \eta_{\star} \sim N_{2n + p}\left(0_{2n + p}, \sigma^2V_{y_\star}\right)\;,
\end{equation}
where $V_{y_\star} = \begin{bmatrix}
        (1 - \rho) I_n & 0 & 0 \\
        0 & M_0 & 0 \\
        0 & 0 & \rho V_\phi
    \end{bmatrix}$
and a prior $\pi(\gamma_\star, \sigma^2, \rho) = \text{IG}(\sigma^2 \given a_{\sigma^2}, b_{\sigma^2}) \times \pi(\rho)$ (since the prior for $\gamma_\star = (\beta^{\T},\gamma^{\T})^{\T}$ has been absorbed into the linear system). By conjugacy of the prior $\pi(\gamma_{\star}, \sigma^2 \given \rho)$ with the Gaussian likelihood in \eqref{eqn:augmented_system},
the resulting conditional posterior joint distribution, $\pi(\gamma_\star, \sigma^2 \given y_{\star}, \rho) \propto \pi(y_{\star} \given \gamma_{\star}, \sigma^2, \rho) \times \pi(\gamma_{\star}, \sigma^2  \given \rho)$, is  
\begin{equation}{\label{eqn:aug_system_post}}
    \pi(\gamma_\star, \sigma^2 \given y, \rho) = \text{N}_{n + p}\left(\gamma_\star \given M_{\star}m_{\star}, \sigma^2M_{\star}\right) \times \text{IG}\left(\sigma^2 \given a_n, b_n\right)\;, 
\end{equation}
where $M_{\star} = (X_\star^{\T}V_{y_\star}^{-1}X_\star)^{-1}$, $m_{\star} = X_\star^{\T}V_{y_\star}^{-1}y_\star = \left(\frac{1}{1 - \rho}y^{\T}X + m_0^{\T}, \frac{1}{1 - \rho}y^{\T}\right)^{\T}$, $a_n = a_{\sigma^2} + \frac{n}{2}$, and $b_n = b_{\sigma^2} + \frac{1}{2}\left(\frac{1}{1 - \rho}y^{\T}y + m_0^{\T}M_0m_0 - m_{\star}^{\T}M_{\star}m_{\star}\right)$. The posterior mean $M_{\star}m_{\star} = \hat{\gamma}_{\star}$ is the general least squares regression estimate to be solved $X_\star^{\T}V_{y_\star}^{-1}X_\star\hat{\gamma}_{\star} = X_\star^{\T}V_{y_\star}^{-1}y_{\star}$ and 
$b_n = b_{\sigma^2} + \frac{ns^2}{2}$, where $s^2 = \frac{1}{n}(y_{\star} - X_{\star}\hat{\gamma}_{\star})^{\T}V_{y_\star}^{-1}(y_{\star} - X_{\star}\hat{\gamma}_{\star})$ is the mean residual sums of squares from \eqref{eqn:augmented_system}.

Fixing $\rho \in (0,1)$ we can draw exact posterior samples from \eqref{eqn:aug_system_post}. 
Since $\gamma = \sigma\sqrt{\rho}\phi$, 
we obtain  
\begin{equation}\label{eqn:phi_post_var}
    \Var(\phi \given y, \sigma^2, \rho) =  \left(V_\phi^{-1} + \frac{\rho}{1 - \rho}\left(I_n - XA^{-1}X^{\T}\right)\right)^{-1},
\end{equation}
where $A = X^{\T}X + (1 - \rho) M_0^{-1}$. A flat prior to $\beta$, that is, $M_0^{-1} = O$, yields $A=X^{\T}X$ and 
\begin{equation} \label{eqn:phi_post_var_flatbeta}
    \lim_{M_0^{-1} \to O} \Var(\phi \given y, \sigma^2, \rho) =  \left(V_\phi^{-1} + \frac{\rho}{1 - \rho}\left(I_n - H\right)\right)^{-1},\;\mbox{ where }\; H = X(X^{\T}X)^{-1}X^{\T}\;.
\end{equation}
Section~\ref{subsec:ME_contrasts} uses these expressions  to formulate the detection of differences between boundaries.

\subsection{Limiting cases}\label{subsec: limits_of_rho}
The limits of the conditional posterior distribution of $(\beta, \gamma)$ as $\rho \rightarrow 0^{+}$ or as $\rho \rightarrow 1^{-}$ are analytically derived when the prior is non-informative on $\beta$. By spectral decomposition, $V^{1/2}_\phi (I - H) V^{1/2}_\phi = CDC^{\T}$ where $C$ is an orthogonal matrix and $D = \text{diag}(d_1,\ldots, d_n)$ with entries that are nonnegative eigenvalues of $V_\phi^{1/2}\left(I_n - H\right)V_\phi^{1/2}$. Let $U = V^{1/2}_\phi C$, then $U$ is invertible, $U^{\T}V^{-1}_\phi U = I_n$, and $U^{\T}(I_n - H)U = D$.
The block entries of $M_{\star}$ and $\hat{\gamma}_{\star}$ in \eqref{eqn:aug_system_post} reveal that as $\rho \rightarrow 0^{+}$,
\begin{equation}\label{eqn:nonspatial_limit_dist}
    \begin{split}
        \Ep[\gamma \given y, \sigma^2, \rho] &= U\left(\frac{1 - \rho}{\rho}I_n + D\right)^{-1}DU^{-1}y \longrightarrow 0_{n}, \\
        \Var(\gamma \given y, \sigma^2, \rho) &= \sigma^2 U\left(\frac{1}{\rho}I_n + \frac{1}{1 - \rho} D \right)^{-1}U^{\T} \longrightarrow 0_{n \times n}.
    \end{split}
\end{equation}
Therefore, with a flat prior on $\beta$,  
$\lim_{\rho \rightarrow 0^+} \pi(\beta, \gamma, \sigma^2 \given y, \rho) = \delta_{0_n}(\gamma)$
(degenerate mass at $\gamma = 0_n$).  

We also obtain
$\lim_{\rho \rightarrow 1^{-}} \left(\frac{1}{\rho}I_n + \frac{1}{1 - \rho} D \right)^{-1} = I_n - D^\star$ and $\lim_{\rho \rightarrow 1^{-}} \left(\frac{1 - \rho}{\rho}I_n + D\right)^{-1}D = D^{\star}$ where $D^\star \in \R^{n\times n}$ is diagonal with $D_{ii}^\star = I(d_i > 0)$ for $i = 1, \ldots, n$.
Simplifications reveal
\begin{equation}\label{eqn:spatial_limit_dist}
        \begin{split}
            \Ep[\beta \given y, \sigma^2, \rho] &\longrightarrow (X^{\T}X)^{-1}X^{\T}\left(y - UD^{\star}U^{-1}e\right), \quad \Ep[\gamma \given y, \sigma^2, \rho] \longrightarrow UD^{\star}U^{-1}e,\\
            \Var(\beta \given y, \sigma^2, \rho) &\longrightarrow \sigma^2(X^{\T}X)^{-1}X^{\T}VX(X^{\T}X)^{-1}, \quad\Var(\gamma \given y, \sigma^2, \rho) \longrightarrow \sigma^2 V,
        \end{split}
\end{equation}
as $\rho \rightarrow 1^{-}$, where $e = (I - H)y$ are the residuals from ordinary least squares and $V = V_\phi - UD^{\star}U^{\T}$.
Finally, $\lim_{\rho \rightarrow 1^{-}} b_n = b_{\sigma^2} + \frac{1}{2}v^{\T}D^\star v$, where $v = U^{-1}e$. 
Therefore, $\lim_{\rho \rightarrow 1^{-}} \pi(\sigma^2 \given y, \rho) = \text{IG}(\sigma^2 \given a_{\sigma^2} + n / 2, b_{\sigma^2} + v^{\T}D^\star v / 2)$. Section~\ref{subsec:epsilon_intro} discusses their implications for boundary detection.
%


\section{A New Framework for Detecting Spatial Disparities}\label{sec: new_methods}

\subsection{Connection to Classical Hypothesis Testing}\label{subsec:connection_to_frequentist}
We first briefly address contrast testing in Bayesian linear regression without spatial effects, that is, $\gamma = 0$ in \eqref{eqn:BYM2_model}. Whereas $\Pr(c_k^{\T}\beta = 0\given y) = 0$ for a continuous posterior distribution, we can test $H^{(k)}_0: c_k^{\T}\beta = 0$ against $H^{(k)}_1: c_k^{\T}\beta \neq 0$ for $K$ nonzero $p\times 1$ vectors $c_1, \ldots, c_K$ using a classical linear model, $y = X\beta + \eta, \eta \sim \mbox{N}_n(0_n, \sigma^2 V_y)$, where $\beta$ is fixed but unknown \citep[e.g.,][]{seberHypothesisTesting2003}. We compute 
$t_{\text{obs}}^{(k)} = \frac{c_k^{\T}\hat{\beta}}{\hat{\sigma}\sqrt{c_{k}^{\T}(X^{\T}V_y^{-1}X)^{-1}c_{k}}}$ for each $c_k$, where $\hat{\beta} = (X^{\T}V_y^{-1}X)^{-1}X^{\T}V_y^{-1}y$ and $(n-p)\hat{\sigma}^2 = \left(y - X\hat{\beta}\right)^{\T}V_y^{-1}\left(y - X\hat{\beta}\right)$ are generalized least squares estimates. We reject $H^{(k)}_0$ if the p-values $p_{k} = \Pr\left(T > \left\vert t^{(k)}_{\text{obs}}\right\vert\right)$, where $T \sim t_{n - p}$, are below a multiplicity adjusted threshold. 

In contrast to assuming strict equality in the point null hypothesis, equivalence tests define a margin around the null value where the effect size is not considered to be practically different from the null value. Bayesian approaches based on this ``region of practical equivalence" (ROPE) follow a similar logic and enable full probabilistic statements about the distribution of the parameter within the ROPE \citep{kruschkeBayesianNewStatistics2018, makowskiIndicesEffectExistence2019}. Here, we follow a similar approach by retaining the continuous conjugate prior for $\beta$ and evaluating the posterior probabilities for the absolute standardized linear combinations between elements of $\beta$ to exceed a threshold $\epsilon$. 
Hence, for all $k = 1, \ldots, K$ and any $\epsilon > 0$, we define 
\begin{equation}\label{eqn:FE_defining_vij}
v_k(\epsilon) = \Pr\left(\left.\frac{\vert c_k^{\T}\beta|}{\sigma\sqrt{c_{k}^{\T}Mc_k}} > \epsilon \;\right|\; y\right).
\end{equation}
We refer to these posterior probabilities as $\epsilon$-difference probabilities, where $\epsilon$ represents a universal threshold or equivalence margin across all $K$ linear combinations indicating the number of posterior standard deviation units that $|c_k^{\T}\beta|$ is away from $0$. If $\beta^{(1)},\ldots,\beta^{(N)}$ and $\sigma^{(1)},\ldots, \sigma^{(N)}$ are $N$ samples drawn from their respective posterior distributions, then we estimate \eqref{eqn:FE_defining_vij} using  
\begin{equation}\label{eqn:vk_MC_estimate}
    \hat{v}_{k}(\epsilon) = \frac{1}{N}\sum_{t = 1}^{N} \text{I}\left(\frac{|c_{k}^{\T}\beta^{(t)}|}{\sigma^{(t)}\sqrt{c_k^{\T}M c_k}} > \epsilon \right)
\end{equation}
for any given $\epsilon$. Adapting \eqref{eqn:augmented_system} to $\gamma=0$ yields $\pi(\beta, \sigma^2 \given y) = \mbox{N}(\beta \given Mm, \sigma^2 M) \times \text{IG}(\sigma^2 \given a, b)$, where $a = a_0 + \frac{n}{2}$, $b = b_0  + \frac{1}{2}\left(y^{\T}V_y^{-1}y + m_0^{\T}M_0m_0 - m^{\T}M m\right)$, $M^{-1} = M_0^{-1} + X^{\T}V_y^{-1}X$ and $m = m_0 + X^{\T}V_y^{-1}y$. We draw exact posterior samples from the Normal-Gamma family. Specific domain applications can provide further information on possible constraints for $\epsilon$ and on defining a clinically significant difference between coefficients $\beta_i$ and $\beta_j$, which, for example, could represent mean results for different treatment groups in a Bayesian Analysis Of Variance (ANOVA). 

The classical p-values and the probabilities $\left\{v_k(\epsilon)\right\}_{k}$ defined in \eqref{eqn:FE_defining_vij} have opposite interpretation: a small value of $p_k$ suggests that $|c_k^{\T}\beta|$ is far from $0$, while small $v_k(\epsilon)$ suggests that $|c_k^{\T}\beta|$ is likely to be close to $0$. In fact, Bayesian inference with a flat prior on $\beta$ amounts to using the classical generalized least squares estimate $\hat{\beta}$. Hence, with a specific prior, the p-values and $\epsilon$-difference probabilities yield identical decisions on contrasts in the following sense.

\theorem\label{proposition:identical_rejection_paths}
Let $c_{1}, \ldots, c_{K} \in \R^{p}$, $X \in \R^{n \times p}$, $y \in \R^n$ and $V_y \in \R^{n \times n}$ be fixed. Under the standard linear model $y \sim \mbox{N}_n(X\beta, \sigma^2 V_y)$, let $p_{k}$ be the p-value to test the null hypothesis $H_0^{(k)}: c_{k}^{\T}\beta = 0$ against $H_{1}^{(k)}: c_{k}^{\T}\beta \neq 0$ for $k = 1, \ldots, K$. Then, $p_{k} > p_{k'}$ if and only if $v_{k}(\epsilon) < v_{k'}(\epsilon)$ for all $k \neq k'$, where $v_{k}(\epsilon) = \Pr\left(\left.\frac{|c_{k}^{\T}\beta|}{\sigma\sqrt{c_k^{\T}M c_k}} > \epsilon \;\right|\; y\right)$ under $\pi(\beta, \sigma^2) \propto \pi(\sigma^2)$
. 
\proof See Section~\ref{append:prop1_proof}.

Theorem~\ref{proposition:identical_rejection_paths} implies that the ascending order of the p-values associated with null hypotheses $H_0^{(k)}: c_{k}^{\T}\beta = 0$ with the two-sided alternative $H_{1}^{(k)}: c_{k}^{\T}\beta \neq 0$ is equivalent to the descending order of the Bayesian $\epsilon$-difference probabilities. Since this result holds for all $\epsilon > 0$ and the classical procedure does not depend on $\epsilon$, the Bayesian rankings are stable with respect to $\epsilon$.  

\subsection{Extension to Spatial Random Effects}\label{subsec:ME_contrasts}

We consider \eqref{eqn:BYM2_model}, where each observation corresponds to one region. We let $V_\phi$ be known, while $\rho$ 
can be either a fixed hyperparameter or a random variable with a prior density. Generalizing the $\epsilon$-difference probabilities in the fixed effects model, we perform posterior inference on random variables $r_{k}(\epsilon) = \text{I}\left(\frac{|c_k^{\T}\gamma_\star|}{\sigma\sqrt{c_k^{\T}M_\star c_k}} > \epsilon \right)$ for any given $\epsilon > 0$. For $k = 1\ldots, K$, we define 
\begin{equation}\label{eqn:hk_in_ME_model}
    h_k(\epsilon; \rho) = \Pr\left(\left.\frac{|c_k^{\T}\gamma_\star|}{\sigma\sqrt{c_k^{\T}M_\star c_k}} > \epsilon \;\right|\; y, \rho\right),
\end{equation}
where $c_{k} \in \R^{n + p}$ is a contrast on $\gamma_\star = (\beta^{\T}, \gamma^{\T})^{\T}$. If $c_k^{\T}\gamma_\star = a_k^{\T}\gamma$ for some $a_k \in \R^{n}$, then $h_k(\epsilon; \rho) = \Pr\left(\left.\frac{|a_k^{\T}\phi|}{\sqrt{a_k^{\T}\Var(\phi \given y, \sigma^2, \rho) a_k}} > \epsilon \;\right|\; y, \rho\right)$ where $\Var(\phi \given y, \sigma^2, \rho)$ is given by \eqref{eqn:phi_post_var}. 

In the context of detecting spatial disparities, consider $K$ pairs of neighboring regions $L = \{(i, j) : i < j, i \sim j\}$, where $i \sim j$ means that the regions $i$ and $j$ are neighbors. If $c_{ij} \in \R^{n + p}$ and $a_{ij} \in \R^{n}$ are such that $c_{ij}^{\T}\gamma_\star = a_{ij}^{\T}\gamma = \gamma_i - \gamma_j$, then the corresponding $\epsilon$-difference probability is $h_{ij}(\epsilon; \rho) = \Pr\left(\left.\frac{|\phi_i - \phi_j|}{\sqrt{a_{ij}^{\T}\Var(\phi \given y, \sigma^2, \rho) a_{ij}}} > \epsilon \;\right|\; y, \rho\right)$. Here, $\epsilon$ represents the equivalence margin for the combined contribution from all latent health effects averaged within each area after adjusting for known risk factors in $X$. 
Using \eqref{eqn:BYM2_model} and \eqref{eqn:augmented_system}, we extend the insights obtained from Theorem~\ref{proposition:identical_rejection_paths} and show that the ranking of conditional posterior probabilities $h_1(\epsilon), \ldots, h_K(\epsilon)$ is stable with respect to $\epsilon$ for any $\rho \in (0, 1)$.

\theorem\label{proposition:BYM2_model_rej_paths} 
Let $\pi(\gamma, \beta, \sigma^2) = \text{N}_n(\gamma \given 0, \sigma^2\rho V_\phi ) \times \text{N}_p(\beta \given M_0 m_0, \sigma^2 M_0) \times \text{IG}(\sigma^2 \given a_{\sigma^2}, b_{\sigma^2})$ be the prior distribution for the model in \eqref{eqn:BYM2_model}. Let $c_{1},\ldots, c_{K} \in \R^{n + p}$ and $X \in \R^{n \times p}$ be fixed. Define $h_k(\epsilon; \rho) = \Pr\left(\left.\frac{|c_k^{\T}\gamma_\star|}{\sigma\sqrt{c_k^{\T}M_\star c_k}} > \epsilon \;\right|\; y, \rho\right)$ for $k=1,\ldots,K$, where $\epsilon > 0$ and $\rho \in (0, 1)$. If there exists $\epsilon_\star > 0$ such that $h_{k}(\epsilon_\star; \rho) < h_{k'}(\epsilon_\star; \rho)$ for any $k \neq k'$, then $h_{k}(\epsilon; \rho) < h_{k'}(\epsilon; \rho)$
.
\proof See Section~\ref{append:prop2_proof}.

If $\pi(\rho)$ denotes the prior for $\rho$, then the marginal posterior distribution of $\gamma_\star$ is analytically inaccessible and we define the $\epsilon$-difference probability for any $ c_k \in \mathbb{R}^{(n+p)\times 1}$ as
\begin{equation}\label{eqn:BYM2_vk_eps_diff}
    v_k(\epsilon) = \Pr\left(\left.\frac{|c_k^{\T}\gamma_\star|}{\sigma\sqrt{c_k^{\T}(X_\star^{\T}V_{y_\star}^{-1}X_\star)^{-1}c_k}} > \epsilon \;\right|\; y \right) = \int_0^1 h_k(\epsilon; \rho)\pi(\rho \given y)d\rho\;.
\end{equation}
Given posterior samples $\gamma_\star^{(1)},\ldots, \gamma_\star^{(N)}$ and $\rho^{(1)},\ldots, \rho^{(N)}$, we estimate \eqref{eqn:BYM2_vk_eps_diff} by
\begin{equation}\label{eqn:ME_vk_MCest}
    \hat{v}_k(\epsilon) = \frac{1}{N} \sum_{t = 1}^N \text{I}\left( \frac{|c_k^{\T}\gamma_\star^{(t)}|}{\sigma^{(t)}\sqrt{c_k^{\T}\left(X_\star^{\T}\left(V_{y_\star}^{(t)}\right)^{-1}X_\star\right)^{-1}c_k}} > \epsilon\right)\; ,
\end{equation}
where $V_{y_\star}^{(t)} = \begin{bmatrix}
        (1 - \rho^{(t)}) I_n & 0 & 0 \\
        0 & M_0 & 0 \\
        0 & 0 & \rho^{(t)} V_\phi
    \end{bmatrix}$. Since $V_{y_\star}^{(t)}$ depends on the sampled value of $\rho^{(t)}$, computing $\Var(\gamma_\star \given y, \sigma^2, \rho)$ for each posterior sample requires inverting an $(n + p) \times (n + p)$ matrix for each posterior sample, which can become cumbersome. When we are only interested in linear combinations of spatial effects $\gamma$, the computation of $\Var(\phi \given y, \sigma^2, \rho)$ is significantly accelerated by utilizing $\Var(\phi \given y, \sigma^2, \rho) = U\left(I_n + \frac{\rho}{1 - \rho}D\right)^{-1}U^{\T}$ as an alternate expression for \eqref{eqn:phi_post_var_flatbeta} in Algorithm~\ref{alg:gibbs_sampling}, where $U$ and $D$ are defined in Section~\ref{subsec: limits_of_rho} with a flat prior on $\beta$.

The choice of $\pi(\rho)$ is crucial for efficiently sampling the posterior joint of $\{\beta, \gamma, \sigma^2, \rho\}$ using MCMC and computing \eqref{eqn:ME_vk_MCest}. Non-informative priors, such as uniform distributions on $[0, 1]$, induce extreme posterior overfitting in overly parameterized models \citep{gelmanPriorDistributionsVariance2006, simpsonPenalisingModelComponent2017}. Instead, we consider the class of penalized complexity (PC) priors introduced in \cite{simpsonPenalisingModelComponent2017} as a means to encourage model parsimony and construct priors imparting a better interpretation. We follow \cite{rieblerIntuitiveBayesianSpatial2016} by placing a PC prior on $\rho$, denoted by $\mbox{PC}(\rho \given \lambda_{\rho}, V_\phi) \propto \lambda_{\rho} \exp(-\lambda_{\rho} d(\rho; V_{\phi}))$ for $0 \leq \rho \leq 1$, where $\lambda_{\rho} > 0$ is a fixed hyperparameter, $d(\rho; V_\phi) = \sqrt{2D_{KL}(p(y; \rho)||q(y))}$ and $D_{KL}(p(y; \rho)||q(y))$ is the Kullback-Leiber divergence between $p(y; \rho) = \text{N}_n(y \given 0_n, \rho V_\phi + (1 - \rho)I_n)$ and the base model without structured variance $q(y) = \text{N}_n(y \given 0_n, I_n)$. The hyperparameter $\lambda_{\rho}$ is chosen such that $P(\rho \leq U) = a$ where $U$ and $a$ are constants. \cite{rieblerIntuitiveBayesianSpatial2016} present promising simulation results demonstrating the PC prior's ability to shrink towards the simpler models with only spatial or non-spatial error variance. When $\pi(\beta, \gamma, \sigma^2, \rho) = \text{N}_n(\gamma \given 0, \sigma^2 \rho V_\phi) \times \text{IG}(\sigma^2 \given a_{\sigma^2}, b_{\sigma^2}) \times \mbox{PC}(\rho\given \lambda_\rho, V_\phi)$, Algorithm \ref{alg:update_steps} provides Gibbs updates for $\{\beta, \gamma, \sigma^2\}$ and a Metropolis-Hastings update for $\rho$. We assess predictive fit using data replicates, where one instance of $y_{rep}^{(t)} \sim \text{N}_n\left(X\beta^{(t)} + \gamma^{(t)}, \sigma^{2(t)}(1 - \rho^{(t)})I_n\right)$ is sampled for each drawing of the posterior samples $\{\beta^{(t)}, \gamma^{(t)}, \sigma^{2(t)}, \rho^{(t)}\}$.

\subsection{Difference Probabilities in Generalized Linear Mixed Models}\label{subsec:glmm_diff_prob}

The stability of rankings of conditional posterior probabilities is established in the previous sections using a normally distributed outcome. More generally, for generalized linear mixed models theoretical tractability is lost when the outcome follows a probability distribution from the exponential family with link function 
\begin{equation}\label{eqn:glmm_model}
    g(\Ep(y_{i} \given \beta, \gamma_{i}, \eta_{i})) = x_{i}^{\T}\beta + \gamma_{i} + \eta_{i},
\end{equation} 
where $x_i$ is $p \times 1$ consisting of values of covariates for the $i$-th observation, $i = 1, \ldots, n$, $\gamma \sim \mbox{N}_{n}(\gamma \given 0_n, \sigma^2 \rho V_{\phi})$, and $\eta \sim \mbox{N}_n(0, \sigma^2(1 - \rho) I_n)$. The link function $g(\cdot)$ relates the mean result to covariates and random effects and is commonly chosen as the canonical link such that \eqref{eqn:glmm_model} is the natural parameter of the exponential family distribution of $y_i$.

We use posterior samples using MCMC to estimate difference probabilities of the form 
\begin{equation}\label{eqn:marginal_var_diff_prob}
    \tau_{k}(\epsilon) = \Pr \left(\left.\frac{|c_k^{\T}\gamma_{\star}|}{\sqrt{\Var(c_k^{\T} \gamma_{\star} \given y)}} > \epsilon \;\right|\; y \right), 
\end{equation}
where $\gamma_{\star} = (\beta^{\T}, \gamma^{\T})^{\T}$. The marginal posterior variance is estimated with the posterior samples by $\hat{\Var}(c_k^{\T} \gamma_{\star} \given y) = \frac{1}{T - 1}\sum_{t = 1}^{T} (c_k^{\T}\gamma_{\star} - c_k^{\T} \bar{\gamma}_{\star})^2$ where $\bar{\gamma}_{\star} = \frac{1}{T} \sum_{t = 1}^T \gamma^{(t)}_{\star}$.
The order of $\tau_1, \ldots, \tau_K$  may still be stable with respect to choice of $\epsilon$ if the marginal posterior distribution of $\gamma_{\star}$ resembles, for example, a multivariate normal or t distribution.

We note that this approach for non-Gaussian outcomes has a weakness for detecting spatial disparities: the choice of the link function $g(x)$ in \eqref{eqn:glmm_model} critically determines the interpretation of the difference probability. For example, if $g(x) = \log(x)$, the canonical link for a Poisson generalized linear model, then the differences $|\phi_{i} - \phi_{j}|$ must now be interpreted in terms of unscaled difference in adjusted log-relative risk, which may produce non-intuitive results. The difference probabilities corresponding to the boundaries in high-risk boundaries may be ranked lower than those of low-risk regions with similar geographic heterogeneity in observed rates. Depending on other risk factors, disparities in log-relative risk may be more frequent in low-risk regions with minimal differences in incidence rates. Difference probabilities based on transformations of the spatial residuals can offer one alternative resolution but may result in a different set of reported boundaries; we demonstrate this through a simulation example in Section \ref{subsec:poisson_sim}, where the difference probability defined in \eqref{eqn:marginal_var_diff_prob} achieves superior classification performance when compared to a similar difference probability defined in terms of an adjusted incidence rate ratio. We otherwise focus our attention on the case of a normally distributed outcome and identity link function, applying equal weight to differences between rates in high-risk and low-risk regions.

\subsection{A Bayesian FDR-Based Decision Rule.}\label{subsec:bayesian_FDR_procedure}

After computing the difference probability $v_{k}(\epsilon)$ for each linear combination of interest, we wish to account for multiple comparisons before constructing binary decisions because we are not operating under an automatically penalizing mixture model as in \cite{scottExplorationAspectsBayesian2006}. We specifically safeguard against Type I errors by controlling the false discovery rate \citep[FDR introduced in][]{benjaminiControllingFalseDiscovery1995}, while retaining power by minimizing the false negative rate (FNR). Thus, we define $\text{FDR} = \frac{\sum_{k = 1}^K \text{I}\left(\frac{\vert c_k^{\T}\gamma_\star|}{\sigma\sqrt{c_{k}^{\T}M_\star c_k}} \leq \epsilon\right)d_{k}}{\iota + \sum_{k = 1}^K d_{k}}$, $
    \text{FNR} = \frac{\sum_{k = 1}^K \text{I}\left(\frac{\vert c_k^{\T}\gamma_\star|}{\sigma\sqrt{c_{k}^{\T}M_\star c_k}} > \epsilon\right)(1 - d_{k})}{\iota + \sum_{k = 1}^K (1 - d_{k})}$
where $\iota > 0$ is a small constant to avoid a zero denominator and $d_{k} \in \{0, 1\}$ represents our decision to declare $\frac{\vert c_k^{\T}\gamma_\star|}{\sigma\sqrt{c_{k}^{\T}M_\star c_k}} > \epsilon$. The FDR and FNR functions are random variables that have different interpretations in the two paradigms \cite{mullerFDRBayesianMultiple2006}. 

Given a set of difference probabilities $\{v_1(\epsilon),\ldots, v_K(\epsilon)\}$, we use the Bayesian FDR and FNR defined in \cite{mullerOptimalSampleSize2004} as the posterior expectation
\begin{equation} \label{eqn:FDR_defn}
    \overline{\text{FDR}}(t^{\star}, \epsilon) = \Ep[\text{FDR}(t^{\star}, \epsilon) \given y] =  \frac{\sum_{k = 1}^K (1 - v_{k}(\epsilon))\text{I}(v_{k}(\epsilon) \geq t^{\star})}{\sum_{k = 1}^K \text{I}(v_{k}(\epsilon) \geq t^{\star})},
\end{equation}
\begin{equation}\label{eqn:FNR_defn}
    \overline{\text{FNR}}(t^{\star}, \epsilon) = \Ep[\text{FNR}(t^{\star}, \epsilon) \given y] =  \frac{\sum_{k = 1}^K v_{k}(\epsilon)\text{I}(v_{k}(\epsilon) < t^{\star})}{\sum_{k = 1}^K \text{I}(v_{k}(\epsilon) < t^{\star})}.
\end{equation}
Minimizing the $\overline{\text{FNR}}$ subject to $\overline{\text{FDR}} \leq \delta$ is achieved by decisions of the form $d_{k} = \text{I}(v_{k}(\epsilon) \geq t^{\star}(\epsilon))$, where $t^{\star}(\epsilon) \in [0, 1]$ for all $\epsilon > 0$ is a threshold chosen as
\begin{equation}\label{eqn:t_cutoff}
    t^{\star}(\epsilon) = \inf \left\{\; t \in [0, 1] : \overline{\text{FDR}}(t, \epsilon) \leq \delta \right\}.
\end{equation}

Bayesian and classical FDR compute posterior probabilities and p-values, respectively, which are used to rank the plausibility of each hypothesis before choosing a cut-off according to a decision rule. In general, the Bayesian FDR and its classical analogue are fundamentally different criterion that are not necessarily equivalent, and their respective control procedures may produce different sets of decisions. However, in the linear fixed effects model and the setting of Theorem~\ref{proposition:identical_rejection_paths}, if the prior is noninformative on $\beta$ in the Bayesian approach, then the order of the posterior probabilities is the reversed order of the p-values. In this case, the Bayesian and classical FDR control procedures differ only in the decision criterion; specific choice of the decision parameter $\delta$ in the Bayesian approach can afford decisions identical to many classical FDR control procedures. 

For creating binary decisions, $\epsilon$ must be chosen to apply this Bayesian FDR control procedure. However, for any $\delta > 0$, if we set $\epsilon$ to be arbitrarily small and $t^\star = 0$, we would declare every decision positive as $d_k = 1$. On the other hand, if we set $\epsilon$ to be arbitrarily large and $t^{\star} = 1$, then we would declare every decision negative as $d_k = 0$. Both cases achieve $\overline{\text{FDR}} \leq \delta$, but do not induce meaningful statements about the data. This motivates a heuristic to choose $\epsilon$ at a reasonable value before applying the Bayesian FDR control procedure to create decisions.

\subsection{Selecting an \texorpdfstring{$\epsilon$}{epsilon}-difference threshold}\label{subsec:epsilon_intro}

In the absence of guidance on the choice of $\epsilon$, we consider two general heuristics to select a reasonable value for $\epsilon$, the first based on a loss function and the relationship between $\epsilon$ and our resulting actions. As $\epsilon$ increases, the event that a contrast exceeds $\epsilon$ standard deviations gains stronger meaning, but has a lower posterior probability. On the other hand, the event that a contrast is less than $\epsilon$ standard deviations becomes less precise but more probable. This motivates the use of a loss function that is optimized with respect to $\epsilon$ by balancing the strength of our affirmative conclusions that a boundary is a disparity at the $\epsilon$-difference level with the precision of our negative conclusions.

As a general heuristic, we appeal to the principle of maximum entropy, which suggests that, given a selection of possible posterior distributions, the posterior distribution that best represents our posterior knowledge is precisely the one with maximum entropy. The principle of maximum entropy \citep{jaynesInformationTheoryStatistical1957} selects the maximum entropy distribution due to it being ``maximally noncommittal" with respect to the amount of information we imbue the posterior with beyond the observed data. Under the model \eqref{eqn:BYM2_model}, choosing an extremely small or large difference threshold $\epsilon$ causes the posterior distribution of $r_{k}(\epsilon) = \text{I}\left(\frac{|c_k^{\T}\gamma_\star|}{\sigma\sqrt{c_k^{\T}M_\star c_k}} > \epsilon \right)$ to collapse into a point mass at 1 or 0 respectively, which has minimal entropy. Maximizing the joint posterior entropy of $R = \left\{r_{1}(\epsilon),\ldots, r_{K}(\epsilon)\right\}$ with respect to $\epsilon$ thus avoids obscuring the information that the data $y$ provides on the true quantities of interest, $\left\{\frac{|c_1^{\T}\gamma_\star|}{\sigma\sqrt{c_1^{\T}M_\star c_1}}, \ldots, \frac{|c_K^{\T}\gamma_\star|}{\sigma\sqrt{c_K^{\T}M_\star c_K}}\right\}$. 

\cite{shannonMathematicalTheoryCommunication1948} explains the Shannon entropy as an objective measure of the information or uncertainty of a probability distribution. Let $I$ be a $K \times 1$ vector such that $I_k \in \{0, 1\}$ for all $k = 1,\ldots, K$ and let $g(I) = \Pr(\cap_{k = 1}^K r_{k}(\epsilon) = I_{k} \given y)$. Since $r_{k}(\epsilon)$ is a binary random variable equal to 1 with posterior probability $v_{k}(\epsilon)$, the joint Shannon posterior entropy of $R$ measured in nats is defined as $H(R) = - \sum_{I \in \{0, 1\}^{K}} g(I) \log g(I)$. However, the joint entropy entails estimating $2^K$ probabilities and is computationally unfeasible even for moderate $K$. 

Instead, we opt for maximizing the entropy of a uniformly chosen $\epsilon$-difference indicator. For any $\epsilon > 0$, let $J$ be a random variable with uniform probability mass on $\{1,\ldots, K\}$ and $W_{\epsilon} = r_J(\epsilon)$. The conditional posterior entropy of $W_{\epsilon}$ given $J$ is $H_J(W_{\epsilon}) 
= \frac{1}{K} \sum_{k = 1}^K H(r_{k}(\epsilon))$. The sum of the individual entropies is related to the joint entropy by the sub-additivity property of the joint entropy \citep{shannonMathematicalTheoryCommunication1948}, which states that $H(R) \leq \sum_{k = 1}^K H(r_{k}(\epsilon))$, where $H(r_k(\epsilon)) = -v_k(\epsilon)\log v_k(\epsilon) - (1 - v_k(\epsilon))\log(1 - v_k(\epsilon))$. 
We obtain an optimal $\epsilon$ by maximizing the average of the individual entropies 
or, equivalently, minimizing 
\begin{equation}\label{eqn:cond_ME_loss}
    \text{LOSS}_{\text{CE}}(\epsilon) = \sum_{k = 1}^K v_{k}(\epsilon)\log v_{k}(\epsilon) + (1 - v_{k}(\epsilon))\log(1  - v_{k}(\epsilon))\;,
\end{equation}
which we refer to as the conditional entropy loss function. We denote the $\epsilon$ obtained from minimizing \eqref{eqn:cond_ME_loss} as $\epsilon_{CE}$.  When $v_{k}(\epsilon)$ is unavailable in closed form, this loss function is estimated from the posterior samples using \eqref{eqn:vk_MC_estimate}. For difference boundary analysis on a set $L$ containing $K$ pairs of neighboring regions and a corresponding set of difference probabilities, $\{v_{ij}(\epsilon)\}_{(i, j) \in L}$, the summation in \eqref{eqn:cond_ME_loss} is over all pairs of neighboring regions, $(i, j) \in L$. In the special case of perfect spatial health equity where all difference probabilities are exactly equal for all $\epsilon$, this procedure results in $v_{(i,j)}(\epsilon_{CE}) = 0.5$ for all boundaries $(i, j)$. Also, $v_k(\epsilon)$ is replaced with $h_k(\epsilon; \rho)$ defined in \eqref{eqn:hk_in_ME_model} when conditioning on a given $0 < \rho < 1$.

As an alternative to optimizing \eqref{eqn:cond_ME_loss}, we can consider the number of positive decisions indicated by $T$. Fixing two quantities out of $\epsilon$, $\delta$ and $T$ determines the third, representing a three-way trade-off between protection against Type I errors, quantity, and quality in our positive results: $\delta$ indicates our tolerance for false positives; $T$ is the quantity of positive results; and $\epsilon$ is our standard for a disparity, which reflects the quality of our decisions. If $\delta$ and $T$ are fixed, 
then a valid choice of $\epsilon$ should belong to $A_\epsilon = \left\{\epsilon > 0 : \overline{\text{FDR}}\left(t^\star(\epsilon)\right) \leq \delta, \; \sum_{(i, j) \in L} d_{ij} \geq T  \right\}$. 
This criterion can be more useful in policy settings aiming to identify rankings of the most egregious disparities, for example, the top fifty. However, this approach will always report $T$ disparities, even if there are no significant differences between the regions. Finally, by fixing $\epsilon$ and $T$, $\delta$ implicitly equals the Bayesian FDR of the resulting decisions, although this does not represent active protection against Type I errors. 

We conclude this section with a remark on the implications of the limiting cases discussed in Section~\ref{subsec: limits_of_rho}. As $\rho$ tends to $0$, \eqref{eqn:nonspatial_limit_dist} indicates that boundary detection becomes meaningless because spatial effects converge to $0$. As $\rho$ tends to 1, $\sigma^2$ and $\gamma$ also have limiting distributions described 
in \eqref{eqn:spatial_limit_dist}, suggesting that the conditional variance of $\gamma$ (in the definition of the $\epsilon$-difference probability) or $\sigma^2$ will neither explode nor shrink toward $0$ when $\rho$ is near 1. 
 
\section{Simulation Experiments}\label{sec:simulation_experiments}

\subsection{Simulation Comparison with ARDP-DAGAR}\label{sec:ARDP_comparison}

We compare difference boundary classification performance between our $\epsilon$-difference approach and the areally-referenced Dirichlet process (ARDP) method introduced by \cite{li2015bayesian} and expanded to the multivariate setting in \cite{gaoSpatialDifferenceBoundary2023} in 100 simulated datasets on a map of the Californian counties. 

\begin{figure}[!t]
    \centering
    \includegraphics[width=0.5\linewidth]{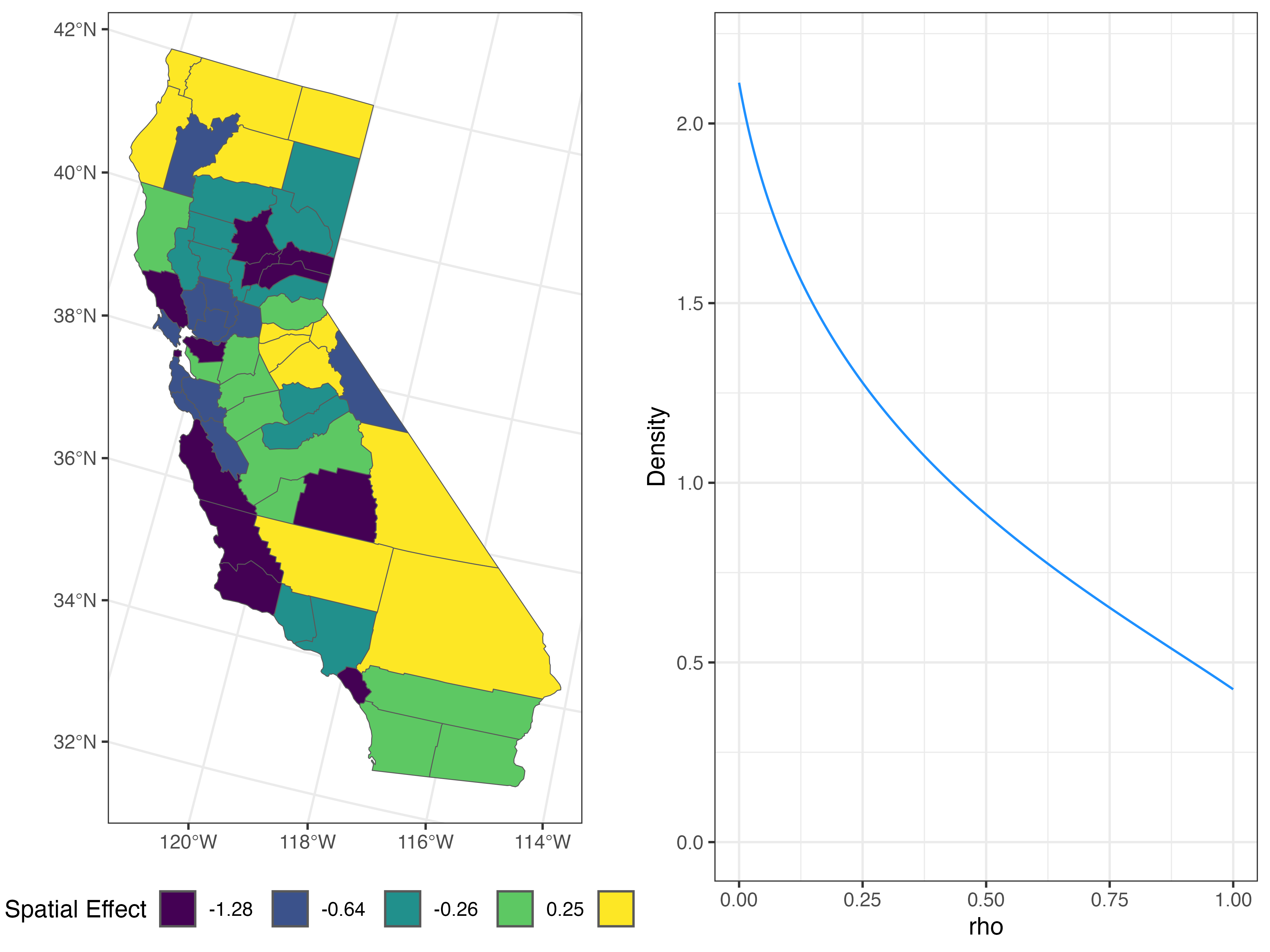}
      \caption{Left: simulated unscaled spatial effects $\phi$. There are 90 difference boundaries separating counties with distinct spatial effects. Right: prior density of $\rho \sim \mbox{PC}(\lambda_{\rho} = 0.2)$.}
    \label{fig:sim_setup}
\end{figure}

To create a reference dataset, we first draw $z \sim \mbox{N}_n(0_n, \Sigma)$, where $\Sigma$ is a exponential covariance kernel such that $\Sigma_{ij} = \exp(-d_{ij} / 150)$ where $d_{ij}$ is the Great Circle distance between the centroids of regions $i$ and $j$ in kilometers. We divide the components $z_1, \ldots, z_n$ into quintiles and replace each component with the mean of the corresponding quintile. This neutral setup allows for a fair comparison between our proposed method and an existing Bayesian non-parametric approach. The resulting spatial effects $\phi$ are fixed across datasets and mapped in the left panel of Figure~\ref{fig:sim_setup}. Out of 139 total, there are 90 boundaries $(i, j)$ where $\phi_i \neq \phi_j$, which we label as the true difference boundaries. We then generate 100 datasets $\{X^{(i)}, y^{(i)}\}$ by generating covariates $X^{(i)} = (1_n, x^{(i)}), x^{(i)} \sim \mbox{N}_{n}(0, I_n)$ and the response $y^{(i)} \sim \mbox{N}_n(X^{(i)}\beta + \gamma, \sigma^2(1 - \rho) I_n)$ with $\gamma = \sqrt{\sigma^2 \rho} \phi$, $\rho = 0.95$, $\sigma^2 = 5$, and $\beta = (2, 5)^{\T}$. 

For both methods, we compute difference probabilities over $L$, the set of all boundaries between Californian counties. For the $\epsilon$-difference method, we employ the BYM2 model in \eqref{eqn:BYM2_model} with the prior $\pi(\beta, \gamma, \sigma^2, \rho) = \mbox{N}_{n}(\gamma \given 0_n, \sigma^2 \rho V_{\phi}) \times \mbox{IG}(\sigma^2 \given 0.001, 0.001) \times \mbox{Unif}(\rho \given 0, 0.99)$. We use a spatial CAR prior with $V_{\phi}^{-1} = c(D_W - \alpha W)$ constructed using the adjacency relations of the Californian counties, $\alpha = 0.99$, and $c = 0.8352$ such that the geometric mean of the marginal prior variances of $\gamma$ is equal to one. We choose $\lambda_{\rho} = 0.2$, corresponding to a prior belief that $\mathbb{P}(\rho \leq 0.5) \approx 0.67$ and truncate the support of $\rho$ to the interval $[0, 0.99]$ to improve convergence. The density of $\mbox{PC}(\rho \given \lambda_{\rho} = 0.2)$ is shown in Figure~\ref{fig:sim_setup}. For each dataset, we draw $20,000$ posterior samples of $(\beta, \gamma, \rho, \sigma^2)$ after 40,000 burn-in samples using the Metropolis within Gibbs sampling algorithm detailed in Appendix~\ref{sec:bym2_sampling_algorithm}, obtain $\epsilon_{CE}$ by minimizing \eqref{eqn:cond_ME_loss}, and estimate the difference probabilities $\{v_{ij}(\epsilon_{CE})\}_{(i, j) \in L}$ using \eqref{eqn:ME_vk_MCest}.

The ARDP model builds spatial dependence through a Markov random field on a latent component and imbues the spatial effects with a probability distribution. Following the notation of \cite{li2015bayesian}, we write the ARDP model as 
\begin{equation}\label{eqn:ARDP}
\begin{split}
    &Y \given \beta, \gamma, \sigma^2 \sim \mbox{N}_{n}(X\beta + \gamma, \sigma^2 I_n), \quad \gamma \sim G_n, \quad G_n = \sum_{u_1, \ldots, u_n} \pi_{u_1, \ldots, u_n} \delta_{\zeta_{u_i}} \cdots \delta_{\zeta_{u_n}}; \\
    &\pi_{u_1, \ldots, u_n} = \Pr\left(\sum_{k = 1}^{u_1 - 1} p_k < F^{(1)}(\phi_1) < \sum_{k = 1}^{u_1} p_k, \ldots, \sum_{k = 1}^{u_n - 1} p_k < F^{(n)}(\phi_n) < \sum_{k = 1}^{u_n} p_k\right); \\
    &\zeta \sim  N_{K}(0, \sigma^2_{s}), \;\; p_1 = V_1, \;\; p_j = V_j \prod_{k < j}(1 - V_k), \;\; V_j \overset{i.i.d}{\sim} \mbox{Beta}(1, \alpha_V), \;\; \phi \sim \mbox{N}_n(0, W_{\phi}), \\
\end{split}
\end{equation}
where $F^{(i)}(\cdot)$ is the CDF of the marginal component $\phi_{i}$, $u_1, \ldots, u_n$ are indices sampled from $\{1, \ldots, N\}$, and $\{p_1, \ldots, p_K\}$ are the stick-breaking weights truncated to $K$ clusters. The spatial components $\phi$ follow a directed acyclic graphical autoregression (DAGAR) model parameterized by $r \in (0, 1)$, a spatial autocorrelation parameter \citep{datta2019spatial}. The DAGAR model requires a specific ordering of the regions $\{1, \ldots, n\}$, which we construct by sorting $\delta_1 < \delta_2 < \ldots < \delta_n$ where $\delta_i = L_{i, 1} - L_{i, 2}$ and $(L_{i, 1}, L_{i, 2})$ are the coordinates of a Albers projection of the $i$th county's centroid. Denote $N(i)$ as the set of neighboring regions that precede $i$ in the ordering and $a_{i}$ is the number of regions in $N(i)$. Then, $W_{\phi}^{-1} = (I_n - B)^{\T}R_{\phi}(I_n - B)$, where $B_{ij} = 0$ if $j \notin N(i)$, $B_{ij} = \frac{r}{1 + (a_i - 1)r^2}$ for $i = 2, \ldots, n$ and $j \in N(i)$, and $R_{\phi}$ is a diagonal matrix with $(R_{\phi})_{ii} = \frac{1 + (a_i - 1)r^2}{1 - r^2}$ for $i = 1, \ldots, n$. 

We match the simulation settings in \cite{gaoSpatialDifferenceBoundary2023} by setting the prior $\pi(\beta, \sigma^2, \sigma^2_s, r) = \mbox{N}_p(\beta \given 0_p, 100^2 I_p) \times \mbox{IG}(\sigma^2 \given 2, 0.1) \times \mbox{IG}(\sigma^2_{s} \given 2, 1) \times \mbox{Unif}(r \given 0, 0.99)$ 
and fixing $\alpha_V = 1$. For each simulated data set, we draw 50,000 $\times$ 4 chains = 200,000 samples from the joint posterior $\pi(\beta, \gamma, \sigma^2, \sigma^2_{s}, \phi, r \given y)$ after 50,000 burn-in samples per chain using the MCMC algorithm implemented in \cite{gaoSpatialDifferenceBoundary2023}. We use the posterior samples of $\gamma$ to compute Monte Carlo estimates of the difference probabilities $w_{ij} = \Pr(\gamma_{i} \neq \gamma_{j} \given y)$ for all neighboring pairs $(i, j) \in L$.

We also compare our results to spatial outlier detection with Local Moran's I computed with ``rook contiguity'' weights based on the standardized residuals $z_i = e_i - \frac{1}{n}\sum_{j} e_j$ from an ordinary least squares regression of $y^{(i)}$ on $x^{(i)}$. We compute significance p-values for each region using 50,000 permutations and control the FDR at $0.15$ with the classical method \citep{benjaminiControllingFalseDiscovery1995}. If a region is significant after multiplicity adjustment, it is a high-low outlier if $z_i > 0, \; \sum_{j\,:\,i \sim j}z_j < 0$ or a low-high outlier if $z_i < 0, \; \sum_{j\,:\,i \sim j}z_j > 0$. We report $(i, j)$ as a difference boundary if the region $i$ is a high-low or low-high outlier and the region $j$ is not significant or is not of the same type of outlier.

\begin{figure}[!t]
    \centering
    \includegraphics[width=0.9\linewidth]{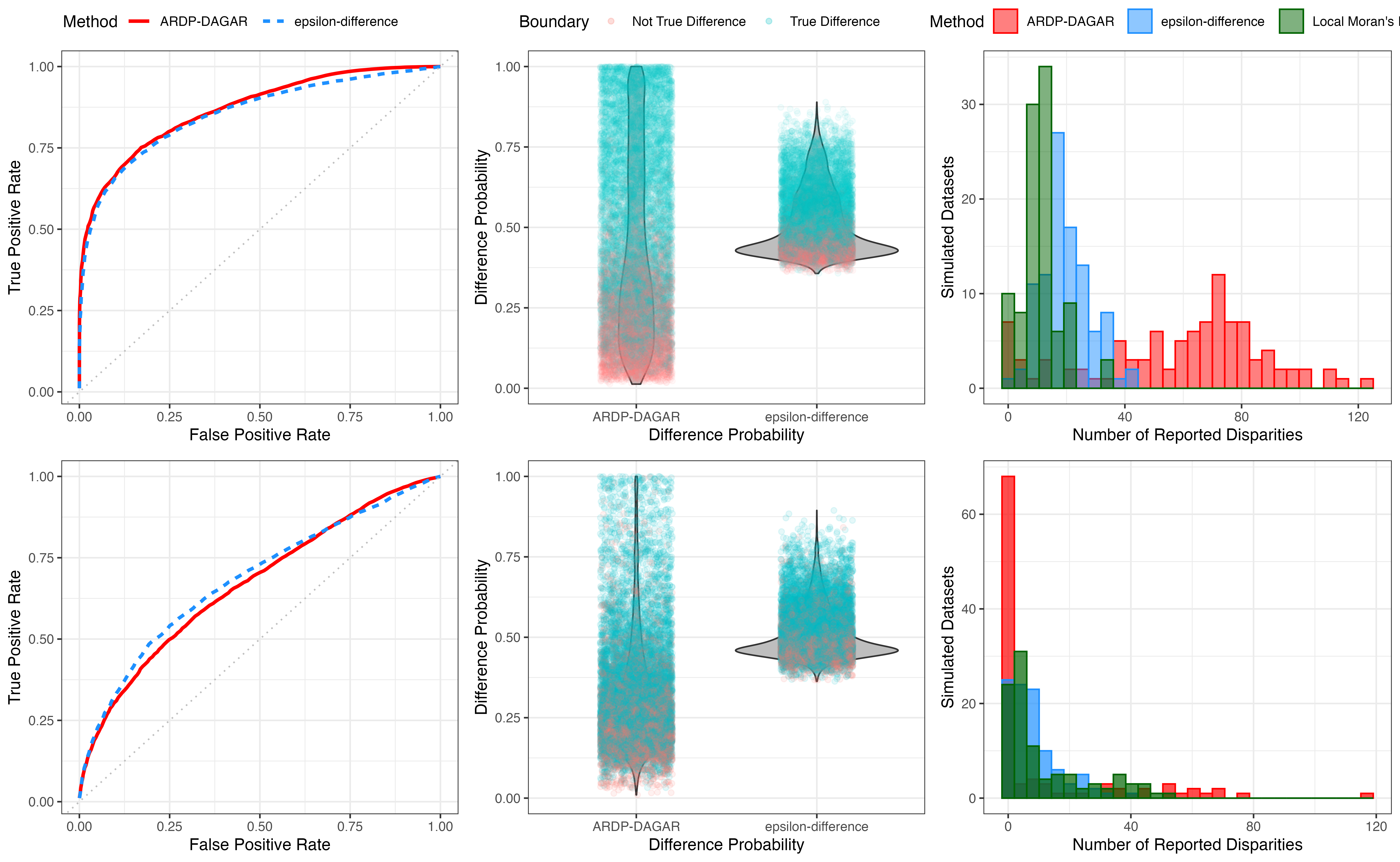}
      \caption{Classification performance of $\epsilon$-difference method with CAR spatial kernel and ARDP-DAGAR method for data simulated with $\rho = 0.95$ (top panels) and $\rho = 0.7$ (bottom panels). Left: ROC curves showing true positive rate versus false positive rate colored by method across varying number of reported difference boundaries. Performance at each cutoff value is averaged across 100 simulated datasets. Middle: violin plots with of difference probabilities generated across each dataset and method. Difference probabilities corresponding to true simulated disparities are colored in blue. Right: histograms of the number of reported disparities in each dataset colored by method. The Bayesian FDR tolerance is set to $\delta = 0.3$ for the ARDP-DAGAR and $\epsilon$-difference methods while the FDR-adjusted p-value significance threshold is set to $0.15$ for boundary detection using Local Moran's I.}
    \label{fig:sim_performance}
\end{figure}

The difference probabilities $v_{ij}(\epsilon)$ and $w_{ij}(\epsilon)$ have equivalent interpretation: a higher difference probability indicates a stronger posterior belief that $(i, j)$ is a true difference boundary. For $T = 1, \ldots, 139$, we take the boundaries corresponding to the top $T$ difference probabilities as the reported difference boundaries in each dataset and average sensitivity and specificity across datasets for each method to obtain the averaged receiver operating characteristic (ROC) curve in the top left panel of Figure~\ref{fig:sim_performance}. For all values of $T$, our proposed framework shows comparable performance to the nonparametric alternative. The area under the ROC curve (AUC) for the $\epsilon$-difference method is 0.856, while the AUC for the ARDP-DAGAR method is 0.871. When $T = 90$, the $\epsilon$-difference method has an averaged sensitivity of 0.828 and a specificity of 0.683, while the ARDP-DAGAR method has an averaged sensitivity of 0.832 and specificity of 0.691. 

We apply the FDR control in Section~\ref{subsec:bayesian_FDR_procedure} with tolerance $\delta = 0.3$ for each data set and both types of difference probabilities. The top difference probabilities from ARDP-DAGAR in the top middle panel of Figure~\ref{fig:sim_performance} belong to a few datasets, resulting in a large variation in the number of reported disparities, as shown in the right panel of Figure~\ref{fig:sim_performance}. The ARDP-DAGAR produced 58.59 (SD = 29.71) disparities reported on average, an overall false discovery rate of 10.68\%, and 6 out of 50 datasets with zero reported disparities. Boundary detection based on Local Moran's I reported 11.87 (SD = 6.63) disparities per data set on average, an overall false discovery rate of 4.30\%, and 10 datasets reporting no disparities. The $\epsilon$-difference approach reported 19.54 (SD = 8.45) disparities on average, an overall false discovery rate of 0.92\%, and zero data sets with no reported disparities. Therefore, our approach shows an attractive decrease in the probability of reporting global non-significance while minimizing the false discovery rate.  

The performance of all three methods suffer greatly when the spatial signal-to-noise ratio decreases. We repeat the simulation runs with $\rho = 0.7$, and re-evaluate the aforementioned performance metrics in the bottom panels of Figure~\ref{fig:sim_performance}. The AUC for the $\epsilon$-difference framework decreases to 0.684 and the AUC for the ARDP-DAGAR method decreases to 0.669. In both cases, our proposed framework on average achieves comparable classification performance to the ARDP-DAGAR approach and results in fewer datasets with zero reported disparities under identical constraints on the Bayesian FDR. Boundary detection using Local Moran's I resulted in a false discovery rate of 18.67\%, while the $\epsilon$-difference (8.97\%) and ARDP-DAGAR (18.42\%) methods provided greater control over the FDR. 

Because of the discrete clustering property of Dirichlet process models, our approach performs better when the spatial residual surface is smooth and spans a wide range of values, while the ARDP-DAGAR method is more suitable when the spatial residuals fall into a small number of distinct levels. Our method may be more efficient for approximately normally distributed outcomes, in particular disease maps in which only one aggregated incidence or prevalence rate is recorded for each areal unit. Compared to spatial outlier detection methods such as those based on Local Indicators of Spatial Association, our approach offers full Bayesian model estimation and FDR control in the presence of non-spatial heterogeneity.

\subsection{Simulated Count Outcome}\label{subsec:poisson_sim}

We demonstrate our analysis framework for a simulated count outcome on a map of the $n = 58$ Californian counties. We simulate a latent risk factor $\gamma \sim \mbox{N}_n(0_n, \sigma^2 \rho W_{\phi})$ and unstructured error $\eta \sim \mbox{N}_n(0_n, \sigma^2 (1 - \rho) I_n)$ with $\sigma^2 = 2, \rho = 0.93$, and $W_{\phi}$ as an exponential covariance kernel such that $W_{\phi, ij} = \exp(-d_{ij} / 50)$, where $d_{ij}$ is the great circle distance between the centroids of county $i$ and $j$ in kilometers. 
We generate covariates $x_{i} = (1, x_{i1})^{\T}, x_{i1} \sim \mbox{N}(2, 1)$ and exposure $E_{i} = \lceil H_{i} \rceil, H_{i} \sim \text{Unif}(10^4, 5 \cdot 10^4)$ for $i = 1, \ldots, 58$. We simulate $y_i \sim \text{Poisson}(E_{i}\exp(x_{i}^{\T}\beta + \gamma_{i} + \eta_{i}))$ with $\beta = (-5, 0.5)^{\T}$. Here, we define the simulated rate as $y_i / E_i$ and denote 
the boundary between county $i$ and $j$ as a true spatial disparity at the $\epsilon$-difference level if $\frac{|\phi_{i} - \phi_j|}{a_{ij}^{\T} W_{\phi}a_{ij}} > \epsilon$.

We analyze the simulated data using a Poisson likelihood with the link function in \eqref{eqn:glmm_model} as $g(y_i) = \log\left(\frac{y_i}{E_{i}}\right)$ and use the prior $\pi(\beta, \gamma, \eta, \sigma^2, \rho) \propto \text{N}_{n}(\gamma \given 0_n, \sigma^2 \rho V_{\phi}) \times \text{N}_{n}(\eta \given 0_n, \sigma^2(1 - \rho)I_n) \times \text{IG}(\sigma^2 \given 0.1, 0.1) \times \mbox{PC}(\rho \given 0.2)$. We place a CAR spatial prior on $\gamma$ by fixing $V_{\phi}^{-1} = c(D_W - \alpha W)$ using the adjacency relations of the Californian counties, $\alpha = 0.99$, and $c = 0.8352$. Using Hamiltonian Monte Carlo implemented within the \texttt{rstan} package in R, we obtain $20,000 \times 4\;\text{chains} = 80,000$ total posterior samples of $(\beta, \gamma, \sigma^2, \rho)$ after discarding 40,000 initial burn-in samples per chain.  We consider $L$, the set of boundaries between two neighboring California counties, and use Monte Carlo estimates of $\{\tau_{ij}(\epsilon)\}_{(i, j) \in L}$ via \eqref{eqn:marginal_var_diff_prob} to minimize the conditional entropy loss and obtain the optimal difference threshold $\epsilon_{CE} = 0.738$. Of 139 total boundaries, 76 are a true disparity at the $\epsilon_{CE} = 0.738$ difference level. 

We repeat the analysis with two alternative difference probability definitions: $\psi_{ij} = \mathbb{P}(|\phi_{i} - \phi_{j}| > \epsilon \given y)$, based on the unstandardized log-rate difference, and $\theta_{ij} = \mathbb{P}(\exp(|\gamma_{i} - \gamma_{j}|) > \epsilon \given y)$, based on the adjusted incidence rate ratio (IRR). For a fair comparison, we also define a pair $(i, j)$ to be a true $\psi$-disparity on the $\epsilon$-difference level if $|\phi_{i} - \phi_{j}| > \epsilon$ or a $\theta$-disparity if $\exp(|\gamma_{i} - \gamma_{j}|) > \epsilon$. We minimize the conditional entropy loss function to obtain $\epsilon_1 = 0.465$ and $\epsilon_2 = 1.506$ for the $\psi_{ij}$ and $\theta_{ij}$ difference thresholds, respectively.

\begin{figure}[!t]
    \centering
    \includegraphics[width=0.9\linewidth]{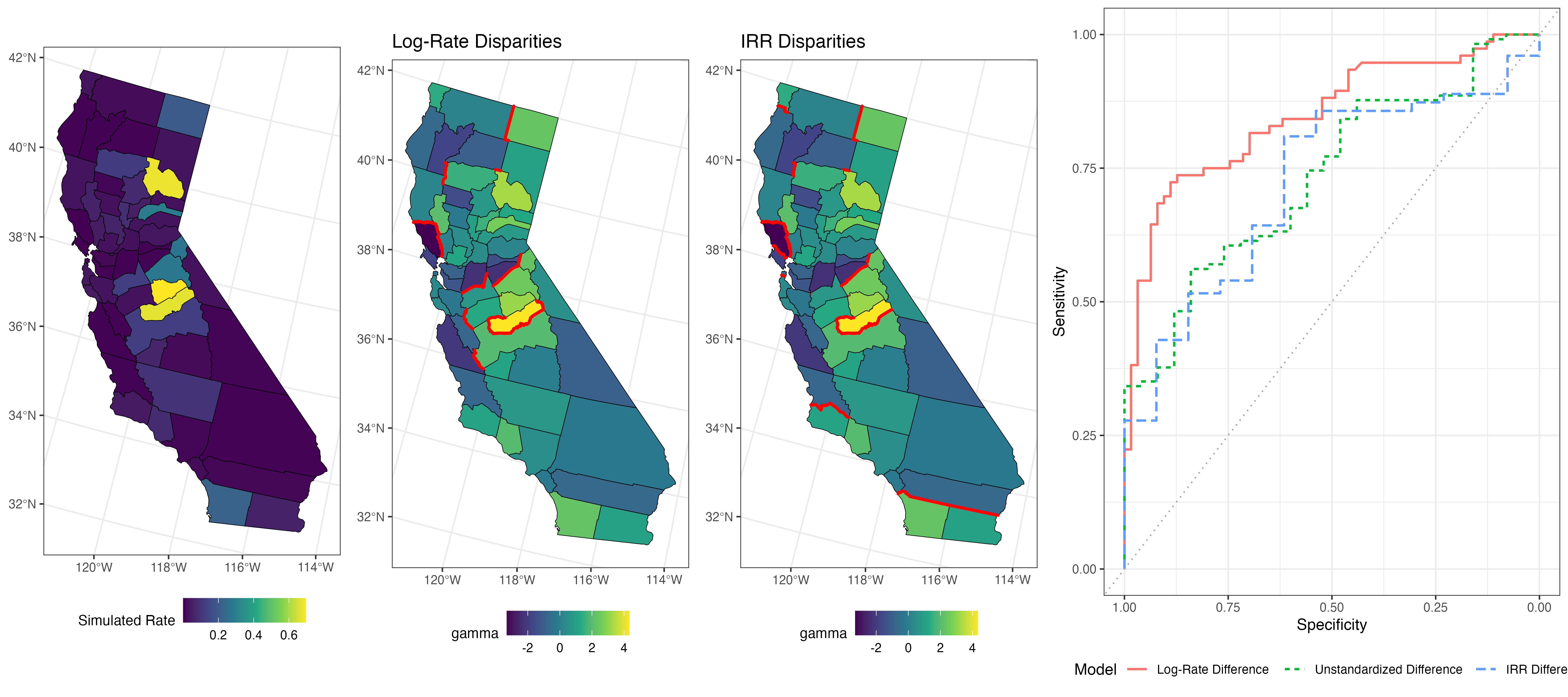}
    \caption{Boundary detection results for simulated count outcome on a California county map. From left to right: (1) Map of simulated rates $y_i / E_i$. (2) Map of simulated latent factors $\gamma$ with (1) boundaries associated with top 20 $\epsilon_{CE} = 0.738$ standardized log-rate difference probabilities $\tau_{ij}$ and (3) boundaries associated with top 20 $\epsilon = 1.506$ IRR difference probabilities $\theta_{ij} = \mathbb{P}(\exp(|\phi_i - \phi_j|) > \epsilon \given y)$. (4) ROC curves showing true positive rate versus true negative rate across varying cutoff values using $\tau_{ij}$ difference probabilities (red; AUC = 0.847), $\psi_{ij} = \mathbb{P}(|\phi_i - \phi_j| > \epsilon \given y)$ difference probabilities (green; AUC = 0.733), and $\theta_{ij} = \mathbb{P}(\exp(|\gamma_i - \gamma_j|) > \epsilon \given y)$ difference probabilities (blue; AUC = 0.723).
    \label{fig:poisson_sim_results}}
\end{figure}

Figure~\ref{fig:poisson_sim_results} shows maps of the simulated rates (left panel) and spatial residuals $\gamma$ with the boundaries corresponding to the top 20 log-rate difference probabilities $\tau_{ij}$ (middle left) or IRR difference probabilities $\theta_{ij}$ (middle right). The two types of difference probabilities share 13 out of the top 20 difference boundaries and both types resulted in all pairwise difference probabilities being below $0.70$. We evaluate the overall classification performance of the difference probabilities $\{v_{ij}(\epsilon)\}_{(i, j) \in L}$ in detecting spatial disparities and obtain a favorable classification performance shown by the ROC curve (AUC = 0.847) in the rightmost panel of Figure~\ref{fig:poisson_sim_results}. We obtain an AUC value of $0.733$ when using $\psi_{ij}$ difference probabilities to detect $\psi$-disparities and an AUC value of $0.723$ when using $\theta_{ij}$ difference probabilities to detect $\theta$-disparities.

To evaluate rank stability, we compute the Spearman correlation between $\{\tau_{ij}(\frac{1}{3}\epsilon_{CE})\}_{(i,j) \in L_S}$ and $\{\tau_{ij}(3\epsilon_{CE})\}_{(i,j) \in L_S}$ as $r_s = 0.903$, which we consider as a rank stability score. The rank stability score using the unstandardized difference in $\psi_{ij}$ is 0.825, indicating that the proposed difference probability formulation $\tau_{ij}$ in \eqref{eqn:marginal_var_diff_prob} offers improved classification performance and robustness in the rankings of the top difference probabilities under this simulation example.

\section{Application}\label{sec:application}

We analyze publicly available health data from the Institute of Health Metrics and Evaluation (IHME) on age-standardized county-level estimates of mortality rates in tracheal, bronchus, and lung cancer in 2014 \citep{mokdadTrendsPatternsDisparities2017}. We adjust for multiple risk factors by incorporating county-level estimates of smoking prevalence in both sexes in 2012, physical inactivity in 2014, percentage of residents 18 years or older without health insurance from 2012 to 2016, unemployment rate percentile (0-1.0) in 2014, overall social vulnerability index (SVI) percentile in 2014, adult diabetes prevalence in 2014, and adult obesity prevalence in 2014. Smoking prevalence was originally derived using data from the Behavioral Risk Factor Surveillance System in \cite{dwyer-lindgrenCigaretteSmokingPrevalence2014}, while data for all other risk factors was downloaded from the publicly available US Diabetes Surveillance System \citep{centersfordiseasecontrolandpreventionUSDiabetesSurveillance}. 
Previous work in \citet{shrevesGeographicPatternsLung2023} indicates that lung cancer mortality rates exhibit spatial autocorrelation that is not completely accounted for by smoking prevalence.

\begin{figure}[!t]
    \centering
    \includegraphics[width=0.9\linewidth]{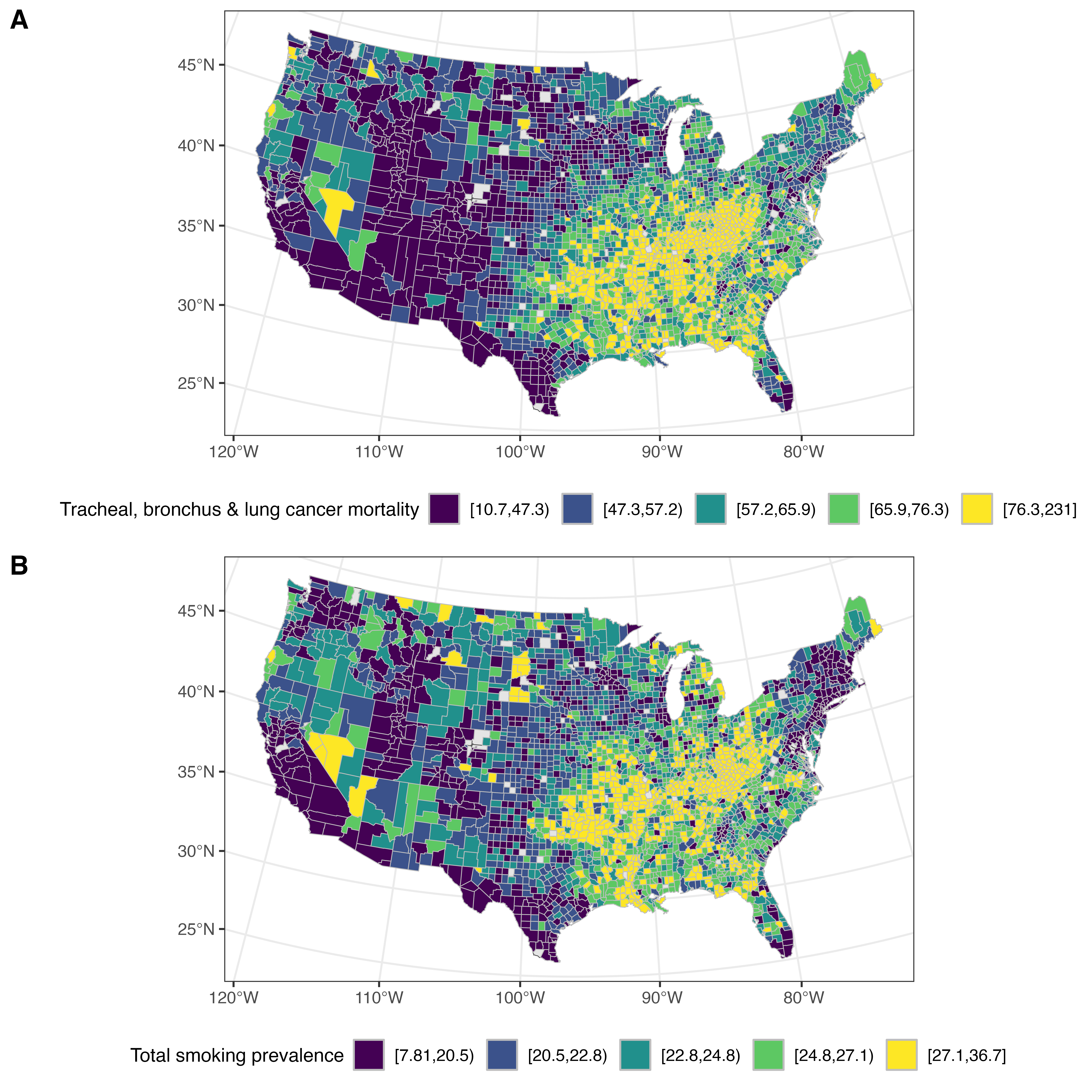}
    \caption{Left: map of county-level age-standardized mortality rate estimates for 
    lung cancer in 2014, colored by quintile. Right: map of county-level total smoking prevalence estimates in 2012, colored by quintile.}
    \label{fig:RDA_maps}
\end{figure}

For our study region, we subset into $n = 3,017$ contiguous US counties with available estimates for smoking prevalence and lung cancer mortality rates with the aim of detecting disparities corresponding to county boundaries. Although we drop counties with missing data for the sake of simplicity in this application, one can refer to \cite{rubinInferenceMissingData1976} for an introduction to Bayesian approaches to missing data or \cite{maBayesianMethodsDealing2018a} for a modern overview. For simplicity, we drop these counties from our study region as the remaining counties form a contiguous region. The cancer mortality rate and estimates of total smoking prevalence for the final study region are shown in Figure~\ref{fig:RDA_maps}. High-risk regions can be clearly identified, most notably in the East South Central states, and our analysis aims to identify local health inequity at the county level. 

To assess spatial autocorrelation, we fit an ordinary least squares linear model with the aforementioned risk factors as predictors of lung cancer mortality before computing Moran's $I = 0.397$ and Geary's $C = 0.586$ using the residuals \citep{banerjeeHierarchicalModelingAnalysis2015}. Both statistics were significant with 10,000 random permutations (p-value = 1/10,001), suggesting that spatial autocorrelation is not fully captured by linearly regressing mortality rates on smoking prevalence. 

We employ \eqref{eqn:BYM2_model} such that $y \given \beta, \gamma, \sigma^2, \rho \sim \text{N}_n\left(X\beta + \gamma, \sigma^2(1 - \rho)I_n\right)$ where $\beta$ is the vector of regression coefficients with an intercept. 
Similarly to the simulation examples, we place a CAR prior on $\phi$ by setting $V_{\phi}^{-1} = c(D_W - \alpha W)$, where the neighbor and adjacency matrices are computed using the subset of US counties, $\alpha = 0.99$, and $c = 0.3762$. We treat $\rho$ as an unknown parameter and specify the joint prior as $\pi(\beta, \gamma, \sigma^2, \rho) =  \text{N}_n(\gamma \given 0, \sigma^2 \rho V_\phi) \times \text{IG}(\sigma^2 \given a_{\sigma^2}, b_{\sigma^2}) \times \text{PC}(\rho \given \lambda_\rho)$, where $a_{\sigma^2} = 0.1$ and $b_{\sigma^2} = 0.1$. Although the set of regions here differs slightly from the simulation example, we also set $\lambda_\rho = 0.0335$ as it approximately satisfies $\Pr(\rho \leq 0.5) = \frac{2}{3}$.

After 10,000 burn-in samples, we draw 30,000 samples from the posterior of $\{\beta, \gamma, \sigma^2, \rho\}$ using Algorithm~\ref{alg:gibbs_sampling}. Table~\ref{tab:RDA_post_summaries} presents 95\% credible intervals for the nonspatial effect parameters, indicating that smoking, SVI percentile, physical inactivity, and uninsured rate are significant predictors of lung cancer mortality. The credible interval of $\rho$ suggests that we are able to learn effectively about $\rho$, and the CAR spatial model explains most of the variance in the data. 

\begin{table}[ht]
\centering
\begin{tabular}{cccc}
  \hline
Parameter & Description & Posterior Mean & 95\% Credible Interval \\ 
  \hline
$\beta_0$ & Intercept & 9.56 & (4.828, 14.223) \\ 
  $\beta_1$ & Smoking prevalence (\%) & 1.90 & (1.754, 2.052) \\ 
  $\beta_2$ & Unemployment percentile & 1.85 & (-0.473, 4.174) \\ 
  $\beta_3$ & SVI percentile & 5.70 & (2.755, 8.68) \\ 
  $\beta_4$ & Physically inactive (\%) & 0.35 & (0.202, 0.49) \\ 
  $\beta_5$ & Uninsured (\%) & -0.60 & (-0.746, -0.451) \\ 
  $\beta_6$ & Diabetes prevalence (\%) & 0.18 & (-0.137, 0.506) \\ 
  $\beta_7$ & Obesity prevalence (\%) & 0.06 & (-0.08, 0.2) \\ 
  $\sigma^2$ & Total variance & 108.25 & (95.927, 123.349) \\ 
  $\rho$ & Spatial proportion of variance & 0.78 & (0.692, 0.861) \\ 
   \hline
\end{tabular}
\caption{Posterior summaries of non-spatial effect parameters in analysis of US county-level lung cancer mortality rates.}
\label{tab:RDA_post_summaries}
\end{table}
We detect spatial disparities using the $\epsilon$-difference framework in Section~\ref{sec: new_methods}. For this dataset, we denote the set of all $K =$~8,793 pairs of neighboring counties as $L = \{(i, j) : i < j, i \sim j \}$ and $\epsilon$-difference probabilities of the form $v_{ij}(\epsilon) = \Pr\left(\left.\frac{|\phi_i - \phi_j|}{\sqrt{a_{ij}^{\T}\Var(\phi \given y, \sigma^2, \rho) a_{ij}}} > \epsilon \;\right|\; y\right)$ for all $(i, j) \in L$. We select the difference threshold $\epsilon$ by minimizing the conditional entropy loss in \eqref{eqn:cond_ME_loss}, which yields $\epsilon_{CE} = 0.922$. For this analysis, we set a maximum allowable Bayesian FDR of $\delta = 0.05$, compute $t^{\star} = 0.884$ using \eqref{eqn:t_cutoff} and report 722 county boundaries exhibiting spatial disparities. 

\begin{figure}[!t]
    \centering
    \includegraphics[width=0.7\linewidth]{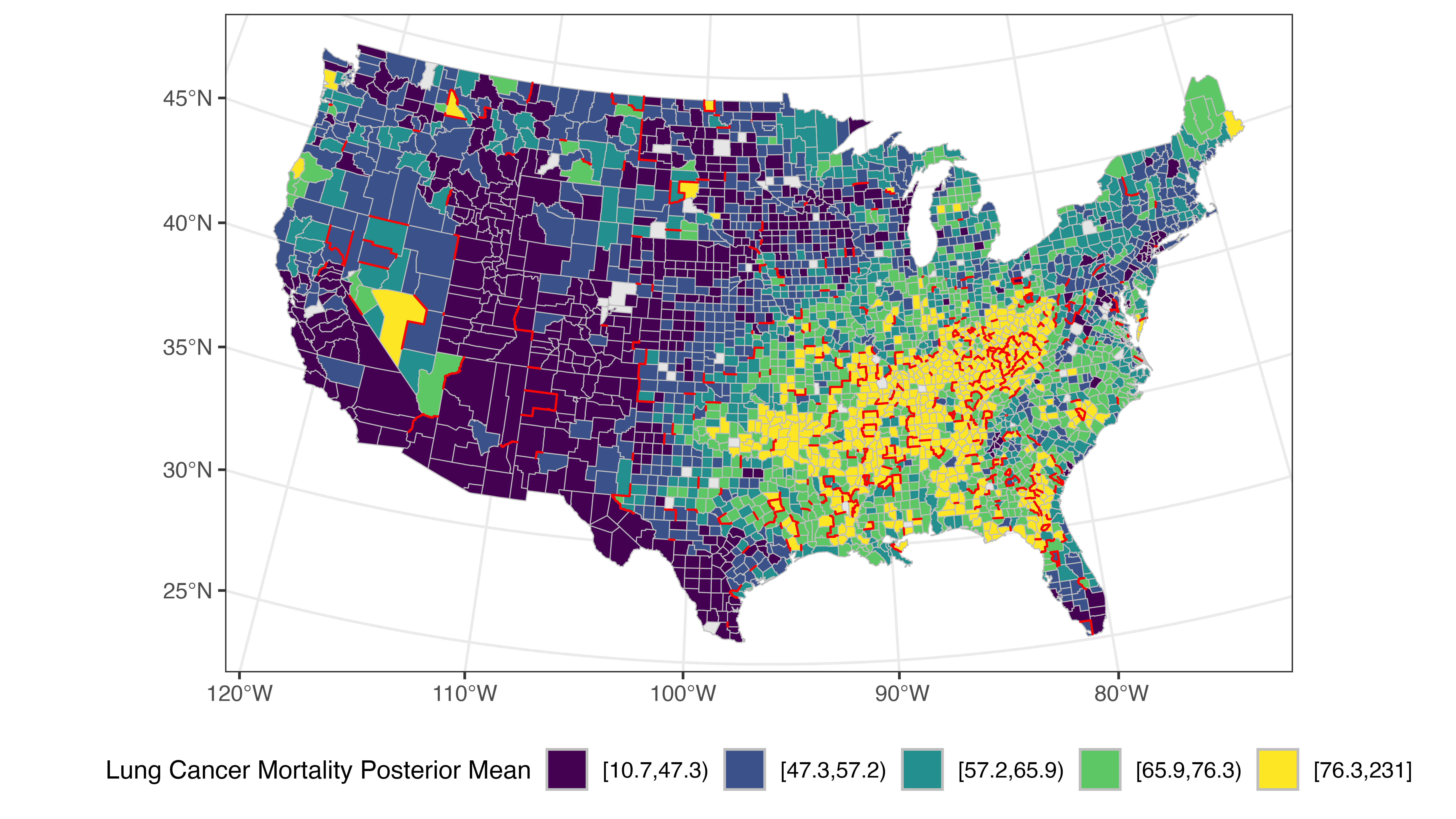}
    \caption{Map of posterior predictive means of 
    lung cancer mortality rates. Red lines correspond to disparities between neighboring counties with difference threshold $\epsilon_{CE} = 0.922$ and Bayesian FDR tolerance $\delta = 0.05$.}
    \label{fig:RDA_postmeans_map}
\end{figure}

 The posterior predictive means and detected disparities are mapped in Figure~\ref{fig:RDA_postmeans_map}. The reported disparities are mainly in relatively high risk states (KY, GA, TN, WV, MS, MS, MO, VA), where several county mortality rate estimates are in the upper quintile with a wide range, suggesting prevalence of health inequities on the local county scale. These disparities between neighboring counties may be caused by a combination of local differences between counties' health policies or other latent determinants of health. Although our model assumes a global effect from each risk factor, previous analysis by \cite{dongVariationFactorsAssociated2022a} suggests that the association of risk factors such as obesity/diabetes prevalence and physical inactivity with cancer mortality may vary by geographical region. 
The complete list of 722 detected difference boundaries ranked by their corresponding difference probabilities is available in Table~\ref{tab:RDA_disparities_table}.

\section{Discussion}\label{sec: discussion}


Our $\epsilon$-based difference boundary detection is considerably simpler to implement and interpret than Bayesian nonparametric approaches \citep{li2015bayesian, gaoSpatialDifferenceBoundary2023}. 
We are able to stabilize the difference probabilities with respect to $\epsilon$, which allows public health officials to report, say, the top 50 disparities between neighboring counties in a robust way. 
We could also define a difference boundary based on the posterior predictive distribution or the posterior distribution of the total residuals $(\gamma + \eta)$ rather than the spatial random effects to identify transitions in the original health outcome or deviations from the explanatory model. We provide some insights in Section~\ref{sec:model_response_diffs} and establish stabilized rankings for the exact conjugate model. This may be preferred in public health practice because of its interpretation in terms of the health outcome itself, but such differences will conflate disparities attributed to space with nonspatial explanatory variables or risk factors. Future areas of investigations will include the impact of spatial confounding on $\epsilon$-based boundary detection, and extending our framework to multivariate modeling with dependent outcomes. For computational efficiency, we will also develop Bayesian predictive stacking \citep[as developed in][for Gaussian process models]{zhangEtAlBayesianPredictiveStacking2025, pan2025bayesianinferencespatialtemporalnongaussian} to avoid MCMC iterations and achieve exact Bayesian boundary detection by stacking over conjugate Bayesian models corresponding to different values of $\rho$.

\section*{Supplementary Materials}

Supplementary materials contain derivation of rank-stability results, sampling algorithm details, formulation of an $\epsilon$-difference probability on the modeled response level, simulation analyses of classification performance sensitivity to the choice of $\epsilon$ and uncertainty in $\rho$, and the full list of spatial disparities in the data analysis. Computer programs that reproduce results in this manuscript are available at \url{https://github.com/Ky-Wu/bayesian_spatial_health_disparities}. 

Algorithms~\ref{alg:gibbs_sampling}~and~\ref{alg:update_steps} are executed in R using \texttt{RcppArmadillo}, the ARDP-DAGAR approach is executed using \texttt{rjags}, and MCMC sampling for the BYM2 Poisson model in Section~\ref{subsec:poisson_sim} is performed using \texttt{rstan}. This work used computational and storage services associated with the Hoffman2 Shared Cluster provided by the UCLA Office of Advanced Research Computing’s Research Technology Group.

\bibliographystyle{abbrvnat}
\bibliography{bayesian_spatial_disparities}

\begin{thebibliography}{57}
\providecommand{\natexlab}[1]{#1}
\providecommand{\url}[1]{\texttt{#1}}
\expandafter\ifx\csname urlstyle\endcsname\relax
  \providecommand{\doi}[1]{doi: #1}\else
  \providecommand{\doi}{doi: \begingroup \urlstyle{rm}\Url}\fi

\bibitem[Aiello and Banerjee(2025)]{aielloDetectingSpatialHealth2025}
L.~Aiello and S.~Banerjee.
\newblock Detecting {{Spatial Health Disparities Using Disease Maps}}.
\newblock https://arxiv.org/abs/2309.02086v2, 2025.

\bibitem[Banerjee et~al.(2015)Banerjee, Carlin, and Gelfand]{banerjeeHierarchicalModelingAnalysis2015}
S.~Banerjee, B.~P. Carlin, and A.~E. Gelfand.
\newblock \emph{Hierarchical {{Modeling}} and {{Analysis}} for {{Spatial Data}}}.
\newblock Number 135 in Monographs on Statistics and Applied Probability. CRC Press, Taylor \& Francis Group, Boca Raton, second edition, 2015.
\newblock ISBN 978-1-4398-1917-3.

\bibitem[{Barboza-Salerno} et~al.(2025){Barboza-Salerno}, Duhaney, Liebhard, and {Schockley-McCarthy}]{barboza-salernoPushingBoundaryChild2025}
G.~E. {Barboza-Salerno}, S.~Duhaney, B.~Liebhard, and K.~{Schockley-McCarthy}.
\newblock Pushing the boundary of child well-being: {{A}} spatial examination of child mortality in transition zones of extreme economic inequality and material hardship.
\newblock \emph{PLOS One}, 20\penalty0 (8):\penalty0 e0330449, Aug. 2025.
\newblock ISSN 1932-6203.
\newblock \doi{10.1371/journal.pone.0330449}.

\bibitem[Benjamini and Hochberg(1995)]{benjaminiControllingFalseDiscovery1995}
Y.~Benjamini and Y.~Hochberg.
\newblock Controlling the {{False Discovery Rate}}: {{A Practical}} and {{Powerful Approach}} to {{Multiple Testing}}.
\newblock \emph{Journal of the Royal Statistical Society: Series B (Methodological)}, 57\penalty0 (1):\penalty0 289--300, 1995.
\newblock ISSN 2517-6161.
\newblock \doi{10.1111/j.2517-6161.1995.tb02031.x}.

\bibitem[Besag et~al.(1991)Besag, York, and Molli{\'e}]{besagBayesianImageRestoration1991}
J.~Besag, J.~York, and A.~Molli{\'e}.
\newblock Bayesian image restoration, with two applications in spatial statistics.
\newblock \emph{Annals of the Institute of Statistical Mathematics}, 43\penalty0 (1):\penalty0 1--20, Mar. 1991.
\newblock ISSN 1572-9052.
\newblock \doi{10.1007/BF00116466}.

\bibitem[Catelan and Biggeri(2010)]{catelanMultipleTestingDisease2010}
D.~Catelan and A.~Biggeri.
\newblock Multiple testing in disease mapping and descriptive epidemiology.
\newblock \emph{Geospatial Health}, 4\penalty0 (2):\penalty0 219--229, May 2010.
\newblock ISSN 1970-7096.
\newblock \doi{10.4081/gh.2010.202}.

\bibitem[{Centers for Disease Control and Prevention}()]{centersfordiseasecontrolandpreventionUSDiabetesSurveillance}
{Centers for Disease Control and Prevention}.
\newblock {{US Diabetes Surveillance System}}.
\newblock https://gis.cdc.gov/grasp/diabetes/diabetesatlas-analysis.html.

\bibitem[Copeland(2010)]{copelandRolePublicHealth2010}
G.~Copeland.
\newblock The role of public health and how boundary analysis can provide a tool for public health investigations: {{The}} public health perspective.
\newblock \emph{Spatial and Spatio-temporal Epidemiology}, 1\penalty0 (4):\penalty0 201--205, Dec. 2010.
\newblock ISSN 1877-5845.
\newblock \doi{10.1016/j.sste.2010.09.002}.

\bibitem[Corpas-Burgos and Martinez-Beneito(2020)]{corpas-burgos2020serra}
F.~Corpas-Burgos and M.~A. Martinez-Beneito.
\newblock On the use of adaptive spatial weight matrices from disease mapping multivariate analyses.
\newblock \emph{Stochastic Environmental Research and Risk Assessment}, 34:\penalty0 531–544, 2020.

\bibitem[Datta et~al.(2019)Datta, Banerjee, Hodges, and Gao]{datta2019spatial}
A.~Datta, S.~Banerjee, J.~S. Hodges, and L.~Gao.
\newblock {Spatial disease mapping using directed acyclic graph auto-regressive (DAGAR) models}.
\newblock \emph{Bayesian analysis}, 14\penalty0 (4):\penalty0 1221, 2019.

\bibitem[Dong et~al.(2022)Dong, Bensken, Kim, Rose, Fan, Schiltz, Berger, and Koroukian]{dongVariationFactorsAssociated2022a}
W.~Dong, W.~P. Bensken, U.~Kim, J.~Rose, Q.~Fan, N.~K. Schiltz, N.~A. Berger, and S.~M. Koroukian.
\newblock Variation in and {{Factors Associated With US County-Level Cancer Mortality}}, 2008-2019.
\newblock \emph{JAMA Network Open}, 5\penalty0 (9):\penalty0 e2230925, Sept. 2022.
\newblock ISSN 2574-3805.
\newblock \doi{10.1001/jamanetworkopen.2022.30925}.

\bibitem[{Dwyer-Lindgren} et~al.(2014){Dwyer-Lindgren}, Mokdad, Srebotnjak, Flaxman, Hansen, and Murray]{dwyer-lindgrenCigaretteSmokingPrevalence2014}
L.~{Dwyer-Lindgren}, A.~H. Mokdad, T.~Srebotnjak, A.~D. Flaxman, G.~M. Hansen, and C.~J. Murray.
\newblock Cigarette smoking prevalence in {{US}} counties: 1996-2012.
\newblock \emph{Population Health Metrics}, 12\penalty0 (1):\penalty0 5, Mar. 2014.
\newblock ISSN 1478-7954.
\newblock \doi{10.1186/1478-7954-12-5}.

\bibitem[Fitzpatrick et~al.(2010)Fitzpatrick, Preisser, Porter, Elkinton, Waller, Carlin, and Ellison]{fitzpatrick2010}
M.~C. Fitzpatrick, E.~L. Preisser, A.~Porter, J.~Elkinton, L.~A. Waller, B.~P. Carlin, and A.~M. Ellison.
\newblock Ecological boundary detection using bayesian areal wombling.
\newblock \emph{Ecology}, 91\penalty0 (12):\penalty0 3448--3455, 2010.
\newblock \doi{https://doi.org/10.1890/10-0807.1}.
\newblock URL \url{https://esajournals.onlinelibrary.wiley.com/doi/abs/10.1890/10-0807.1}.

\bibitem[Gao et~al.(2023)Gao, Banerjee, and Ritz]{gaoSpatialDifferenceBoundary2023}
L.~Gao, S.~Banerjee, and B.~Ritz.
\newblock Spatial {{Difference Boundary Detection}} for {{Multiple Outcomes Using Bayesian Disease Mapping}}.
\newblock \emph{Biostatistics}, 24\penalty0 (4):\penalty0 922--944, Oct. 2023.
\newblock ISSN 1465-4644.
\newblock \doi{10.1093/biostatistics/kxac013}.

\bibitem[Gelman(2006)]{gelmanPriorDistributionsVariance2006}
A.~Gelman.
\newblock Prior distributions for variance parameters in hierarchical models (comment on article by {{Browne}} and {{Draper}}).
\newblock \emph{Bayesian Analysis}, 1\penalty0 (3):\penalty0 515--534, Sept. 2006.
\newblock ISSN 1936-0975, 1931-6690.
\newblock \doi{10.1214/06-BA117A}.

\bibitem[Glickman et~al.(2014)Glickman, Rao, and Schultz]{glickmanFalseDiscoveryRate2014}
M.~E. Glickman, S.~R. Rao, and M.~R. Schultz.
\newblock False discovery rate control is a recommended alternative to~{{Bonferroni-type}} adjustments in health studies.
\newblock \emph{Journal of Clinical Epidemiology}, 67\penalty0 (8):\penalty0 850--857, Aug. 2014.
\newblock ISSN 0895-4356.
\newblock \doi{10.1016/j.jclinepi.2014.03.012}.

\bibitem[Hanson et~al.(2015)Hanson, Banerjee, Li, and McBean]{hanson2015spatial}
T.~Hanson, S.~Banerjee, P.~Li, and A.~McBean.
\newblock Spatial boundary detection for areal counts.
\newblock In \emph{Nonparametric Bayesian Inference in Biostatistics}, pages 377--399. Springer, 2015.

\bibitem[Jacquez(2010)]{jacquezGeographicBoundaryAnalysis2010}
G.~M. Jacquez.
\newblock Geographic boundary analysis in spatial and spatio-temporal epidemiology: {{Perspective}} and prospects.
\newblock \emph{Spatial and Spatio-temporal Epidemiology}, 1\penalty0 (4):\penalty0 207--218, Dec. 2010.
\newblock ISSN 1877-5845.
\newblock \doi{10.1016/j.sste.2010.09.003}.

\bibitem[Jacquez and Greiling(2003{\natexlab{a}})]{jacquez2003a}
G.~M. Jacquez and D.~A. Greiling.
\newblock Geographic boundaries in breast, lung and colorectal cancers in relation to exposure to air toxics in long island, new york.
\newblock \emph{International Journal of Health Geographics}, 2\penalty0 (1):\penalty0 1--22, 2003{\natexlab{a}}.
\newblock b.

\bibitem[Jacquez and Greiling(2003{\natexlab{b}})]{jacquez2003b}
G.~M. Jacquez and D.~A. Greiling.
\newblock Local clustering in breast, lung and colorectal cancer in long island, new york.
\newblock \emph{International Journal of Health Geographics}, 2\penalty0 (1):\penalty0 1--12, 2003{\natexlab{b}}.
\newblock a.

\bibitem[Jagai et~al.(2017)Jagai, Messer, Rappazzo, Gray, Grabich, and Lobdell]{jagai2017county}
J.~S. Jagai, L.~C. Messer, K.~M. Rappazzo, C.~L. Gray, S.~C. Grabich, and D.~T. Lobdell.
\newblock {County-level cumulative environmental quality associated with cancer incidence}.
\newblock \emph{Cancer}, 123\penalty0 (15):\penalty0 2901--2908, 2017.

\bibitem[Jaynes(1957)]{jaynesInformationTheoryStatistical1957}
E.~T. Jaynes.
\newblock Information {{Theory}} and {{Statistical Mechanics}}.
\newblock \emph{Physical Review}, 106\penalty0 (4):\penalty0 620--630, May 1957.
\newblock ISSN 0031-899X.
\newblock \doi{10.1103/PhysRev.106.620}.

\bibitem[Kruschke and Liddell(2018)]{kruschkeBayesianNewStatistics2018}
J.~K. Kruschke and T.~M. Liddell.
\newblock The {{Bayesian New Statistics}}: {{Hypothesis}} testing, estimation, meta-analysis, and power analysis from a {{Bayesian}} perspective.
\newblock \emph{Psychonomic Bulletin \& Review}, 25\penalty0 (1):\penalty0 178--206, Feb. 2018.
\newblock ISSN 1069-9384, 1531-5320.
\newblock \doi{10.3758/s13423-016-1221-4}.

\bibitem[Lawson(2013)]{lawson2013statistical}
A.~B. Lawson.
\newblock \emph{Statistical methods in spatial epidemiology.}
\newblock John Wiley \& Sons, 2013.

\bibitem[Lawson et~al.(2016)Lawson, Banerjee, Haining, and Ugarte]{lawson2016handbook}
B.~Lawson, Andrew, S.~Banerjee, R.~Haining, and D.~Ugarte, Maria.
\newblock \emph{Handbook of Spatial Epidemiology}.
\newblock CRC press, Boca Raton, FL, 2016.

\bibitem[Lee and Mitchell(2012)]{leeBoundaryDetectionDisease2012}
D.~Lee and R.~Mitchell.
\newblock Boundary detection in disease mapping studies.
\newblock \emph{Biostatistics}, 13\penalty0 (3):\penalty0 415--426, July 2012.
\newblock ISSN 1465-4644.
\newblock \doi{10.1093/biostatistics/kxr036}.

\bibitem[Li et~al.(2011)Li, Banerjee, and McBean]{li2011mining}
P.~Li, S.~Banerjee, and A.~M. McBean.
\newblock Mining boundary effects in areally referenced spatial data using the bayesian information criterion.
\newblock \emph{Geoinformatica}, 15\penalty0 (3):\penalty0 435--454, 2011.

\bibitem[Li et~al.(2012)Li, Banerjee, Carlin, and McBean]{banerjee2012bayesian}
P.~Li, S.~Banerjee, B.~P. Carlin, and A.~M. McBean.
\newblock Bayesian areal wombling using false discovery rates.
\newblock \emph{Statistics and its Interface}, 5\penalty0 (2):\penalty0 149--158, 2012.

\bibitem[Li et~al.(2015)Li, Banerjee, Hanson, and McBean]{li2015bayesian}
P.~Li, S.~Banerjee, T.~A. Hanson, and A.~M. McBean.
\newblock Bayesian models for detecting difference boundaries in areal data.
\newblock \emph{Statistica Sinica}, 25\penalty0 (1):\penalty0 385, 2015.

\bibitem[Lu and Carlin(2005)]{lu2005bayesian}
H.~Lu and B.~P. Carlin.
\newblock Bayesian areal wombling for geographical boundary analysis.
\newblock \emph{Geographical Analysis}, 37\penalty0 (3):\penalty0 265--285, 2005.

\bibitem[Lu et~al.(2007)Lu, Reilly, Banerjee, and Carlin]{lu2007bayesian}
H.~Lu, C.~S. Reilly, S.~Banerjee, and B.~P. Carlin.
\newblock Bayesian areal wombling via adjacency modeling.
\newblock \emph{Environmental and ecological statistics}, 14:\penalty0 433--452, 2007.

\bibitem[Ma and Carlin(2007)]{ma2007bayesian}
H.~Ma and B.~P. Carlin.
\newblock Bayesian multivariate areal wombling for multiple disease boundary analysis.
\newblock \emph{Bayesian Analysis}, 2\penalty0 (2):\penalty0 281--302, 2007.

\bibitem[Ma et~al.(2010)Ma, Carlin, and Banerjee]{ma2010hierarchical}
H.~Ma, B.~P. Carlin, and S.~Banerjee.
\newblock Hierarchical and joint site-edge methods for medicare hospice service region boundary analysis.
\newblock \emph{Biometrics}, 66\penalty0 (2):\penalty0 355--364, 2010.

\bibitem[Ma and Chen(2018)]{maBayesianMethodsDealing2018a}
Z.~Ma and G.~Chen.
\newblock Bayesian methods for dealing with missing data problems.
\newblock \emph{Journal of the Korean Statistical Society}, 47\penalty0 (3):\penalty0 297--313, Sept. 2018.
\newblock ISSN 2005-2863.
\newblock \doi{10.1016/j.jkss.2018.03.002}.

\bibitem[Makowski et~al.(2019)Makowski, {Ben-Shachar}, Chen, and L{\"u}decke]{makowskiIndicesEffectExistence2019}
D.~Makowski, M.~S. {Ben-Shachar}, S.~H.~A. Chen, and D.~L{\"u}decke.
\newblock Indices of {{Effect Existence}} and {{Significance}} in the {{Bayesian Framework}}.
\newblock 10, Dec. 2019.
\newblock \doi{10.3389/fpsyg.2019.02767}.

\bibitem[Mokdad et~al.(2017)Mokdad, {Dwyer-Lindgren}, Fitzmaurice, Stubbs, {Bertozzi-Villa}, Morozoff, Charara, Allen, Naghavi, and Murray]{mokdadTrendsPatternsDisparities2017}
A.~H. Mokdad, L.~{Dwyer-Lindgren}, C.~Fitzmaurice, R.~W. Stubbs, A.~{Bertozzi-Villa}, C.~Morozoff, R.~Charara, C.~Allen, M.~Naghavi, and C.~J.~L. Murray.
\newblock Trends and {{Patterns}} of {{Disparities}} in {{Cancer Mortality Among US Counties}}, 1980-2014.
\newblock \emph{JAMA}, 317\penalty0 (4):\penalty0 388--406, Jan. 2017.
\newblock ISSN 0098-7484.
\newblock \doi{10.1001/jama.2016.20324}.

\bibitem[M{\"u}ller et~al.(2004)M{\"u}ller, Parmigiani, Robert, and Rousseau]{mullerOptimalSampleSize2004}
P.~M{\"u}ller, G.~Parmigiani, C.~Robert, and J.~Rousseau.
\newblock Optimal {{Sample Size}} for {{Multiple Testing}}.
\newblock \emph{Journal of the American Statistical Association}, 99\penalty0 (468):\penalty0 990--1001, Dec. 2004.
\newblock ISSN 0162-1459.
\newblock \doi{10.1198/016214504000001646}.

\bibitem[M{\"u}ller et~al.(2007)M{\"u}ller, Parmigiani, and Rice]{mullerFDRBayesianMultiple2006}
P.~M{\"u}ller, G.~Parmigiani, and K.~Rice.
\newblock {{FDR}} and {{Bayesian Multiple Comparisons Rules}}.
\newblock In \emph{Bayesian Statistics 8: {{Proceedings}} of the Eighth Valencia International Meeting June 2--6, 2006}. Oxford University Press, July 2007.
\newblock ISBN 978-0-19-921465-5.
\newblock \doi{10.1093/oso/9780199214655.003.0014}.

\bibitem[Pan et~al.(2025)Pan, Zhang, Bradley, and Banerjee]{pan2025bayesianinferencespatialtemporalnongaussian}
S.~Pan, L.~Zhang, J.~R. Bradley, and S.~Banerjee.
\newblock Bayesian inference for spatial-temporal non-{G}aussian data using predictive stacking.
\newblock 2025.
\newblock URL \url{https://arxiv.org/abs/2406.04655}.

\bibitem[Rao(2023)]{rao2023}
J.~S. Rao.
\newblock \emph{Statistical Methods in Health Disparity Research}.
\newblock Chapman \& Hall/CRC, Boca Raton, FL, 2023.

\bibitem[Riebler et~al.(2016)Riebler, S{\o}rbye, Simpson, and Rue]{rieblerIntuitiveBayesianSpatial2016}
A.~Riebler, S.~H. S{\o}rbye, D.~Simpson, and H.~Rue.
\newblock An intuitive {{Bayesian}} spatial model for disease mapping that accounts for scaling.
\newblock \emph{Statistical Methods in Medical Research}, 25\penalty0 (4):\penalty0 1145--1165, Aug. 2016.
\newblock ISSN 0962-2802.
\newblock \doi{10.1177/0962280216660421}.

\bibitem[Rodriguez et~al.(2018)Rodriguez, Hu, Kershaw, Hastings, L{\'o}pez, Cullen, Harrington, and Palaniappan]{rodriguez2018county}
F.~Rodriguez, J.~Hu, K.~Kershaw, K.~G. Hastings, L.~L{\'o}pez, M.~R. Cullen, R.~A. Harrington, and L.~P. Palaniappan.
\newblock {County‐Level Hispanic Ethnic Density and Cardiovascular Disease Mortality}.
\newblock \emph{Journal of the American Heart Association}, 7\penalty0 (19):\penalty0 e009107, 2018.

\bibitem[Rubin(1976)]{rubinInferenceMissingData1976}
D.~B. Rubin.
\newblock Inference and missing data.
\newblock \emph{Biometrika}, 63\penalty0 (3):\penalty0 581--592, Dec. 1976.
\newblock ISSN 0006-3444.
\newblock \doi{10.1093/biomet/63.3.581}.

\bibitem[Scott and Berger(2006)]{scottExplorationAspectsBayesian2006}
J.~G. Scott and J.~O. Berger.
\newblock An exploration of aspects of {{Bayesian}} multiple testing.
\newblock \emph{Journal of Statistical Planning and Inference}, 136\penalty0 (7):\penalty0 2144--2162, July 2006.
\newblock ISSN 0378-3758.
\newblock \doi{10.1016/j.jspi.2005.08.031}.

\bibitem[Seber and Lee(2003)]{seberHypothesisTesting2003}
G.~A. Seber and A.~J. Lee.
\newblock Hypothesis {{Testing}}.
\newblock In \emph{Linear {{Regression Analysis}}}, chapter~4, pages 97--118. John Wiley \& Sons, Ltd, 2003.
\newblock ISBN 978-0-471-72219-9.
\newblock \doi{10.1002/9780471722199.ch4}.

\bibitem[Shannon(1948)]{shannonMathematicalTheoryCommunication1948}
C.~E. Shannon.
\newblock A {{Mathematical Theory}} of {{Communication}}.
\newblock \emph{Bell System Technical Journal}, 27\penalty0 (3):\penalty0 379--423, July 1948.
\newblock ISSN 00058580.
\newblock \doi{10.1002/j.1538-7305.1948.tb01338.x}.

\bibitem[Shreves et~al.(2023)Shreves, Buller, Chase, Creutzfeldt, Fisher, Graubard, Hoover, Silverman, Devesa, and Jones]{shrevesGeographicPatternsLung2023}
A.~H. Shreves, I.~D. Buller, E.~Chase, H.~Creutzfeldt, J.~A. Fisher, B.~I. Graubard, R.~N. Hoover, D.~T. Silverman, S.~S. Devesa, and R.~R. Jones.
\newblock Geographic {{Patterns}} in {{U}}.{{S}}. {{Lung Cancer Mortality}} and {{Cigarette Smoking}}.
\newblock \emph{Cancer Epidemiology, Biomarkers \& Prevention}, 32\penalty0 (2):\penalty0 193--201, Feb. 2023.
\newblock ISSN 1055-9965.
\newblock \doi{10.1158/1055-9965.EPI-22-0253}.

\bibitem[Simpson et~al.(2017)Simpson, Rue, Riebler, Martins, and S{\o}rbye]{simpsonPenalisingModelComponent2017}
D.~Simpson, H.~Rue, A.~Riebler, T.~G. Martins, and S.~H. S{\o}rbye.
\newblock Penalising {{Model Component Complexity}}: {{A Principled}}, {{Practical Approach}} to {{Constructing Priors}}.
\newblock \emph{Statistical Science}, 32\penalty0 (1):\penalty0 1--28, 2017.
\newblock ISSN 0883-4237.

\bibitem[Slack et~al.(2014)Slack, Myers, Martin, and Heymsfield]{slackGeographicConcentrationUs2014}
T.~Slack, C.~A. Myers, C.~K. Martin, and S.~B. Heymsfield.
\newblock The geographic concentration of us adult obesity prevalence and associated social, economic, and environmental factors.
\newblock \emph{Obesity}, 22\penalty0 (3):\penalty0 868--874, Mar. 2014.
\newblock ISSN 1930-7381, 1930-739X.
\newblock \doi{10.1002/oby.20502}.

\bibitem[Slutske et~al.(2023)Slutske, Conner, Kirsch, Smith, Piasecki, Johnson, McCarthy, Nez~Henderson, and Fiore]{slutskeExplainingCOVID19Related2023}
W.~S. Slutske, K.~L. Conner, J.~A. Kirsch, S.~S. Smith, T.~M. Piasecki, A.~L. Johnson, D.~E. McCarthy, P.~Nez~Henderson, and M.~C. Fiore.
\newblock Explaining {{COVID-19}} related mortality disparities in {{American Indians}} and {{Alaska Natives}}.
\newblock \emph{Scientific Reports}, 13\penalty0 (1):\penalty0 20974, Nov. 2023.
\newblock ISSN 2045-2322.
\newblock \doi{10.1038/s41598-023-48260-9}.

\bibitem[Sun et~al.(2015)Sun, Reich, Cai, Guindani, and Schwartzman]{sunFalseDiscoveryControl2015}
W.~Sun, B.~J. Reich, T.~T. Cai, M.~Guindani, and A.~Schwartzman.
\newblock False {{Discovery Control}} in {{Large-Scale Spatial Multiple Testing}}.
\newblock \emph{Journal of the Royal Statistical Society. Series B, Statistical methodology}, 77\penalty0 (1):\penalty0 59--83, Jan. 2015.
\newblock ISSN 1369-7412.
\newblock \doi{10.1111/rssb.12064}.

\bibitem[Tansey et~al.(2018)Tansey, Koyejo, Poldrack, and Scott]{tanseyFalseDiscoveryRate2018}
W.~Tansey, O.~Koyejo, R.~A. Poldrack, and J.~G. Scott.
\newblock False {{Discovery Rate Smoothing}}.
\newblock \emph{Journal of the American Statistical Association}, 113\penalty0 (523):\penalty0 1156--1171, July 2018.
\newblock ISSN 0162-1459.
\newblock \doi{10.1080/01621459.2017.1319838}.

\bibitem[Tian et~al.(2010)Tian, Goovaerts, Zhan, and Wilson]{tianIdentificationRacialDisparities2010}
N.~Tian, P.~Goovaerts, F.~B. Zhan, and J.~G. Wilson.
\newblock Identification of racial disparities in breast cancer mortality: Does scale matter?
\newblock \emph{International Journal of Health Geographics}, 9\penalty0 (1):\penalty0 35, July 2010.
\newblock ISSN 1476-072X.
\newblock \doi{10.1186/1476-072X-9-35}.

\bibitem[Ventrucci et~al.(2011)Ventrucci, Scott, and Cocchi]{ventrucciMultipleTestingStandardized2011}
M.~Ventrucci, E.~M. Scott, and D.~Cocchi.
\newblock Multiple testing on standardized mortality ratios: A {{Bayesian}} hierarchical model for {{FDR}} estimation.
\newblock \emph{Biostatistics}, 12\penalty0 (1):\penalty0 51--67, Jan. 2011.
\newblock ISSN 1465-4644.
\newblock \doi{10.1093/biostatistics/kxq040}.

\bibitem[Waller and Carlin(2010)]{waller2010handbook}
L.~Waller and B.~Carlin.
\newblock Disease mapping.
\newblock In A.~E. Gelfand, P.~Diggle, P.~Guttorp, and M.~Fuentes, editors, \emph{Handbook Of Spatial Statistics}, page 217–243. CRC Press, Boca Raton, FL, 2010.

\bibitem[Waller and Gotway(2004)]{waller2004applied}
L.~A. Waller and C.~A. Gotway.
\newblock \emph{Applied spatial statistics for public health data}, volume 368.
\newblock John Wiley \& Sons, 2004.

\bibitem[Zhang et~al.(in press)Zhang, Tang, and Banerjee]{zhangEtAlBayesianPredictiveStacking2025}
L.~Zhang, W.~Tang, and S.~Banerjee.
\newblock Bayesian geostatistics using predictive stacking.
\newblock \emph{Journal of the American Statistical Association}, pages 1--19, in press.
\newblock \doi{10.1080/01621459.2025.2566449}.
\newblock URL \url{https://doi.org/10.1080/01621459.2025.2566449}.

\end{thebibliography}


\begin{thebibliography}{0}
\providecommand{\natexlab}[1]{#1}
\providecommand{\url}[1]{\texttt{#1}}
\expandafter\ifx\csname urlstyle\endcsname\relax
  \providecommand{\doi}[1]{doi: #1}\else
  \providecommand{\doi}{doi: \begingroup \urlstyle{rm}\Url}\fi

\end{thebibliography}

\end{document}


\maketitle

\section{Proof of Theorem~\ref{proposition:identical_rejection_paths}}\label{append:prop1_proof}


Under the standard linear model described in Theorem~\ref{proposition:identical_rejection_paths}, the p-values, $p_{k} = \Pr\left(t > |t^{(k)}_{\text{obs}}|\right)$, are characterized by a common t-distribution CDF evaluated at the observed t-statistics. As CDFs are monotonic functions, $p_k > p_{k'}$ if and only if $\left\vert t^{(k)}_{\text{obs}}\right\vert  < \left\vert t^{(k')}_{\text{obs}}\right\vert$.  Next, it is well known that when $\pi(\beta, \sigma^2) \propto \pi(\sigma^2)$, then $\beta \given y, \sigma^2 \sim N\left(\hat{\beta}, \sigma^2(X^{\T}V_y^{-1}X)^{-1}\right)$, where $\hat{\beta} = (X^{\T}V_y^{-1}X)^{-1}X^{\T}V_y^{-1} y$ is the generalized least-squares estimate. We have that for any $\epsilon > 0$, $\sigma^2 > 0$,
\begin{align*}
    &\Pr\left(\left.\frac{|c_{k}^{\T}\beta|}{\sigma \sqrt{c_k^{\T}M c_k}} > \epsilon \;\right|\; y, \sigma^2 \right) =
    1 - \Pr\left(\left.-\epsilon \leq \frac{c_{k}^{\T}\beta}{\sigma\sqrt{c_k^{\T}(X^{\T}V_y^{-1}X)^{-1}c_k}} \leq \epsilon \;\right|\; y, \sigma^2 \right) \\
    &\qquad = \Phi\left(-\epsilon + \frac{c_k^{\T}\hat{\beta}}{\sigma\sqrt{c_k^{\T}(X^{\T}V_y^{-1}X)^{-1}c_k}}\right) + \Phi\left(-\epsilon - \frac{c_k^{\T}\hat{\beta}}{\sigma\sqrt{c_k^{\T}(X^{\T}V_y^{-1}X)^{-1}c_k}}\right)  \\
    &\qquad\qquad = \Phi\left(\epsilon - \frac{\hat{\sigma}}{\sigma}t^{(k)}_{\text{obs}}\right) + \Phi\left(\epsilon + \frac{\hat{\sigma}}{\sigma}t^{(k)}_{\text{obs}}\right)
\end{align*}
where $\Phi$ denotes the standard normal CDF. We now use the following lemma.

\lemma\label{lemma:one} Let $\epsilon > 0$, $c > 0$, $G: \R \rightarrow [0, 1]$ be a function such that $G'(x)$ is symmetric around zero and strictly decreasing in $|x|$, and let $h: \R \rightarrow [0, 1]$ be given by $h(x) = G(\epsilon - cx) + G(\epsilon + cx)$ for all $x \in \mathbb{R}$. Then, $h$ is symmetric and if $x, y \in \mathbb{R}$, $h(x) > h(y)$ if and only if $|x| > |y|$.

\proof By assumption, $h'(x) = cG'(\epsilon + cx) - cG'(\epsilon - cx)$. Since $G'(x)$ is strictly decreasing in $|x|$, then if $x < 0$, $|\epsilon + cx| < |\epsilon - cx|$ and thus $h'(x) > 0$. If $x > 0$, then $|\epsilon + cx| > |\epsilon - cx|$, so $h'(x) < 0$. Furthermore, $h(-x) = G(\epsilon + cx) + G(\epsilon - cx) = h(x)$, so $h$ is symmetric and strictly increases in $|x|$, which proves the claim. 

By Lemma \ref{lemma:one}, since the standard normal PDF $\Phi'(x)$ is symmetric around zero and strictly decreasing in $|x|$ and $\hat{\sigma} / \sigma > 0$, then
\begin{equation*}
    \Pr\left(\left.\frac{|c_{k}^{\T}\beta|}{\sigma\sqrt{c_k^{\T}M c_k}} > \epsilon \;\right|\; y, \sigma^2 \right) <  \Pr\left(\left.\frac{|c_{k'}^{\T}\beta|}{\sigma\sqrt{c_{k'}^{\T}Mc_{k'}}} > \epsilon \;\right|\; y, \sigma^2 \right) \iff \left\vert t^{(k)}_{\text{obs}}\right\vert < \left\vert t^{(k')}_{\text{obs}}\right\vert
\end{equation*}
for any $\epsilon > 0, \sigma^2 > 0$. If $\left\vert t^{(k)}_{\text{obs}} \right\vert < \left\vert t^{(k')}_{\text{obs}} \right\vert$, integrating out $\sigma^2$ preserves the order as
\begin{multline*}
    v_{k}(\epsilon) = \Pr\left(\left.\frac{|c_{k}^{\T}\beta|}{\sigma\sqrt{c_k^{\T}M c_k}} > \epsilon \;\right|\; y\right) =  \int_{0}^\infty \Pr\left(\left.\frac{|c_{k}^{\T}\beta|}{\sigma\sqrt{c_k^{\T}M c_k}} > \epsilon \;\right|\; y, \sigma^2 \right)\pi(\sigma^2 \given y) d\sigma^2 \\
    <  \int_{0}^\infty \Pr\left(\left.\frac{|c_{k'}^{\T}\beta|}{\sigma\sqrt{c_{k'}^{\T}Mc_{k'}}} > \epsilon \;\right|\; y, \sigma^2 \right)\pi(\sigma^2 \given y) d\sigma^2 = \Pr\left(\left.\frac{|c_{k'}^{\T}\beta|}{\sigma\sqrt{c_{k'}^{\T}Mc_{k'}}} > \epsilon  \;\right|\; y\right) = v_{k'}(\epsilon)
\end{multline*}
which shows that $\left\vert t^{(k)}_{\text{obs}} \right\vert < \left\vert t^{(k')}_{\text{obs}} \right\vert$ implies $v_k(\epsilon) < v_{k'}(\epsilon)$ for all $\epsilon > 0$. If $v_k(\epsilon) < v_{k'}(\epsilon)$ for all $\epsilon > 0$, then this implies that there exists $\epsilon > 0, \sigma^2 > 0$ such that $\Pr\left(\left.\frac{|c_{k}^{\T}\beta|}{\sigma\sqrt{c_k^{\T}M c_k}} > \epsilon \;\right|\; y, \sigma^2 \right) <  \Pr\left(\left.\frac{|c_{k'}^{\T}\beta|}{\sigma\sqrt{c_{k'}^{\T}Mc_{k'}}} > \epsilon \;\right|\; y, \sigma^2 \right)$, which implies $\left\vert t^{(k)}_{\text{obs}}\right\vert < \left\vert t^{(k')}_{\text{obs}}\right\vert$ by the above argument. Therefore,   
\begin{equation*}
    p_{k} > p_{k'} \iff \left\vert t_{\text{obs}}^{(k)}\right\vert < \left\vert t_{\text{obs}}^{(k')}\right\vert \iff \Pr\left(\left.\frac{|c_{k}^{\T}\beta|}{\sigma\sqrt{c_k^{\T}M c_k}} > \epsilon \;\right|\; y, \sigma^2 \right) < \Pr\left(\left.\frac{|c_{k'}^{\T}\beta|}{\sigma \sqrt{c_{k'}^{\T}Mc_{k'}}} > \epsilon \;\right|\; y, \sigma^2 \right) 
\end{equation*}
for all $\sigma^2 > 0$ and $\epsilon >0$. The above inequality holds if and only if $v_{k}(\epsilon) < v_{k'}(\epsilon) \;\forall \epsilon > 0$. 

\section{Proof of Theorem~\ref{proposition:BYM2_model_rej_paths}.}\label{append:prop2_proof}

The posterior conditional distributions for \eqref{eqn:augmented_system} are given by \eqref{eqn:aug_system_post}. Thus, we have $\gamma_\star \given y, \rho, \sigma^2 \sim N_{n + p}\left(M_{\star}m_{\star}, \sigma^2M_{\star}\right)$. For all $k = 1, \ldots, K$, denote $\alpha_k = \frac{c_k^{\T}M_{\star} m_{\star}}{\sqrt{c_k^{\T}(X_\star^{\T}V_{y_\star}^{-1}X_\star)^{-1}c_k}}$. Then, 
\begin{align*}
    \Pr\left(\left.\frac{|c_k^{\T}\gamma_\star|}{\sigma\sqrt{c_k^{\T}(X_\star^{\T}V_{y_\star}^{-1}X_\star)^{-1}c_k}} > \epsilon \;\right|\; y, \rho, \sigma^2 \right) &= 1 - \Pr\left(\left.-\epsilon \leq \frac{c_k^{\T}\gamma_\star}{\sigma\sqrt{c_k^{\T}(X_\star^{\T}V_{y_\star}^{-1}X_\star)^{-1}c_k}} \leq \epsilon \;\right|\; y, \rho, \sigma^2\right) \\
    &= \Phi\left(-\epsilon + \frac{\alpha_k}{\sigma}\right) + \Phi\left(-\epsilon - \frac{\alpha_k}{\sigma}\right)\;.
\end{align*}
By Lemma \ref{lemma:one}, for any $k \neq k'$, $|\alpha_k| < |\alpha_{k'}|$ if and only if $\Pr\left(\left.\frac{|c_k^{\T}\gamma_\star|}{\sigma\sqrt{c_k^{\T}(X_\star^{\T}V_{y_\star}^{-1}X_\star)^{-1}c_k}} > \epsilon \;\right|\; y, \rho, \sigma^2 \right) < \Pr\left(\left.\frac{|c_{k'}^{\T}\gamma_\star|}{\sigma\sqrt{c_{k'}^{\T}(X_\star^{\T}V_{y_\star}^{-1}X_\star)^{-1}c_{k'}}} > \epsilon \;\right|\; y, \rho, \sigma^2 \right)$ for all $\sigma^2 > 0, \epsilon > 0$.  If there exists $\epsilon_\star > 0$ such that $h_k(\epsilon_\star; \rho) < h_{k'}(\epsilon_\star; \rho)$, then there exists $\sigma^2 > 0$ such that $\Pr\left(\left.\frac{|c_k^{\T}\gamma_\star|}{\sigma\sqrt{c_k^{\T}(X_\star^{\T}V_{y_\star}^{-1}X_\star)^{-1}c_k}} > \epsilon_\star \;\right|\; y, \rho, \sigma^2 \right) < \Pr\left(\left.\frac{|c_{k'}^{\T}\gamma_\star|}{\sigma\sqrt{c_{k'}^{\T}(X_\star^{\T}V_{y_\star}^{-1}X_\star)^{-1}c_{k'}}} > \epsilon_\star \;\right|\; y, \rho, \sigma^2 \right)$, which implies $|\alpha_k| < |\alpha_{k'}|$. Thus, for any $\epsilon > 0$,
\begin{align*}
    h_k(\epsilon; \rho) &= \int_{0}^\infty \Pr\left(\left.\frac{|c_k^{\T}\gamma_\star|}{\sigma\sqrt{c_k^{\T}(X_\star^{\T}V_{y_\star}^{-1}X_\star)^{-1}c_k}} > \epsilon \;\right|\; y, \rho, \sigma^2 \right) \pi(\sigma^2\given y, \rho) d\sigma^2 \\
    &< \int_{0}^\infty \Pr\left(\left.\frac{|c_{k'}^{\T}\gamma_\star|}{\sigma\sqrt{c_{k'}^{\T}(X_\star^{\T}V_{y_\star}^{-1}X_\star)^{-1}c_{k'}}} > \epsilon \;\right|\; y, \rho, \sigma^2 \right) \pi(\sigma^2\given y, \rho) d\sigma^2 = h_{k'}(\epsilon; \rho)\;,
\end{align*}
which completes the proof.

\section{Metropolis within Gibbs Sampling Algorithm for BYM2 Model}\label{sec:bym2_sampling_algorithm}

We present a Metropolis-Hastings within Gibbs sampling algorithm for obtaining posterior samples of $\{\beta, \gamma, \sigma^2, \rho\}$ from the model in \eqref{eqn:BYM2_model} using the prior distribution 
\[
\pi(\beta, \gamma, \sigma^2, \rho) = \text{N}_n(\gamma \given 0, \sigma^2 \rho V_\phi) \times \text{IG}(\sigma^2 \given a_{\sigma^2}, b_{\sigma^2}) \times \mbox{PC}(\rho\given \lambda_\rho, V_\phi).
\]
In Algorithm~\ref{alg:update_steps}, $\texttt{chol}$ returns the upper triangular Cholesky factor of a positive definite matrix, \texttt{updateBeta} and \texttt{updateGamma} return $\beta$ and $\gamma$ from $\beta \given y, \gamma, \sigma^2, \rho \sim \text{N}_p\left(F_{\beta}f_{\beta}, \sigma^2F_{\beta}\right)$, where $F_{\beta}^{-1} = X^{T}X/(1-\rho)$ and $f_{\beta} = X^{\T}(y-\gamma)/(1-\rho)$ and $\gamma \given y, \beta, \sigma^2, \rho \sim \text{N}_{n}\left(F_{\gamma}f_{\gamma}, \sigma^2F_{\gamma}\right)$, where 
$F_{\gamma}^{-1} = \frac{1}{1 - \rho} I_n + \frac{1}{\rho} V_{\phi}^{-1}$ and $f_{\gamma} = (y - X\beta)/(1-\rho)$, respectively, %
and \texttt{updateSigmaSq} samples $\sigma^2 \given y, \beta, \gamma, \rho \sim \text{InvGamma}\left(a_{\sigma^2} + \frac{n}{2}, b_{\sigma^2} + \frac{||y - X\beta - \gamma||^2}{2(1 - \rho)}\right)$.

\RestyleAlgo{ruled}
\begin{algorithm}[H]
\caption{Update steps for Metropolis-Hastings within Gibbs sampling of posterior of $(\beta, \gamma, \sigma^2, \rho)$ in BYM2 Model with flat $\beta$ prior and PC $\rho$ prior}
\SetKwFunction{updateBeta}{updateBeta}
\SetKwFunction{updateGamma}{updateGamma}
\SetKwFunction{updateSigmaSq}{updateSigmaSq}
\SetKwFunction{MHupdateRho}{MHupdateRho}
\SetKwProg{Fn}{function}{:}{}
\SetKwFunction{}{}

Given $y, X, R = \texttt{chol}(X^{\T}X), V_\phi^{-1}, m_1 = X^{\T}y$, $P$, $\Lambda = \mbox{diag}(\lambda_1, \ldots, \lambda_n), a_{\sigma^2}, b_{\sigma^2}, \lambda_\rho$:

\Fn{\updateBeta{$\gamma$, $\sigma^2$, $\rho$}}{
    \For{$i = 1, \dots p$} {
        Sample $z_i \sim \mbox{N}(0, \sigma^2(1 - \rho))$.\\
    }
    Set $v_p \leftarrow \texttt{forwardsolve}(R^{\T}, m_1 - X^{\T}\gamma)$;\quad $\beta \leftarrow \texttt{backsolve}(R, v_p + z)$. \\
    \KwRet $\beta$.
}

\Fn{\updateGamma{$\beta$, $\sigma^2$, $\rho$}}{
    Set $v_n \leftarrow y - X\beta$.\\
    Update $v_n \leftarrow \mbox{diag}\left(\frac{\rho}{\rho + (1 - \rho)\lambda_i}\right) P^{\T} v_n$. \\
    \For{$i = 1, \dots n$} {
        Sample $z_i \sim \mbox{N}(0, \sigma^2)$.\\
    }
    Set $\gamma \leftarrow P\left(v_n + \mbox{diag}\left(\sqrt{\frac{\rho(1 - \rho)}{\rho + \lambda_i (1 - \rho}}\right)z\right)$.\\
    \KwRet $\gamma$.
}

\Fn{\updateSigmaSq{$\rho$, $r^2$}}{
     Sample $\tau^2 \sim \Gamma\left(a_{\sigma^2} + \frac{n}{2}, b_{\sigma^2} + \frac{r^2}{2(1 - \rho)}  \right)$;\quad set $\sigma^2 \leftarrow 1 / \tau^2$. \\
     \KwRet $\sigma^2$.
}

\Fn{\MHupdateRho{$\gamma, \sigma^2, \rho, r^2$}}{
    Initialize $\rho_\star \leftarrow 1$. \\
    \While{$\rho_\star \geq 1$}{
        Sample $\tau^2 \sim \Gamma\left(\frac{n}{2} - 1, \frac{r^2}{2\sigma^{2}}\right)$; \quad set $\rho_\star \leftarrow 1 - 1 / \tau^2$.
    }
    Set $c_1 \leftarrow \sqrt{\sum_{i = 1}^n \left(\frac{\rho}{\lambda_i} - \log \left(\frac{\rho}{\lambda_i} + (1 - \rho)\right)\right) - n\rho}$. \\
    Set $c_2 \leftarrow \sqrt{\sum_{i = 1}^n \left(\frac{\rho_\star}{\lambda_i} - \log \left(\frac{\rho_\star}{\lambda_i} + (1 - \rho_\star)\right)\right) - n\rho_\star}$. \\
    Set $\alpha_{MH} \leftarrow \frac{n}{2}\log\left(\frac{\rho}{\rho_\star}\right) + \lambda_\rho \left(c_1 - c_2\right) + \frac{\gamma^{\T}V_\phi^{-1}\gamma}{2\sigma^{2}}\left(\frac{1}{\rho} - \frac{1}{\rho_\star}\right)$. \\
    Sample $u \sim \text{Unif}(0, 1)$.\\
    \DontPrintSemicolon
    \lIf{$\log(u) \leq \alpha_{MH}$}{\KwRet $\rho_\star$.}
    \lElse{\KwRet $\rho$.}
}
\label{alg:update_steps}
\end{algorithm}

\texttt{MHupdateRho} updates $\rho$ using a Metropolis-Hastings random walk by proposing $1 - \rho_{\star} \sim \text{TruncInvGamma}\left(\frac{n}{2} - 1, \frac{||y - X\beta - \gamma||^2}{2\sigma^2}, [0, 1]\right)$, where we denote $Z_1 \sim \text{TruncInvGamma}\left(a, b, [c_L, c_U]\right)$, $a > 0$, $b > 0$, $0 \leq c_L < c_U$ if $\Pr(Z_1 \leq z) = \Pr(Z_2 \leq z \given c_L \leq Z_2 \leq c_U)$ with $Z_2 \sim \text{IG}(a, b)$. The Metropolis-Hastings acceptance probability is equal to $\min\{1, \exp(\alpha_{MH})\}$, where 
\[
\exp(\alpha_{MH}) = \frac{\pi(\beta, \gamma, \sigma^2, \rho_{\star} \given y)Q(\rho_{\star}, \rho)}{\pi(\beta, \gamma, \sigma^2, \rho \given y) Q(\rho, \rho_{\star})},
\]
$\alpha_{MH}$ is the acceptance log-ratio computed in Algorithm~\ref{alg:update_steps}, and $Q(\rho, \rho_{\star})$ is the proposal density for a new candidate $\rho_{\star}$. Expanding via Bayes' Rule yields 
\[
\exp(\alpha_{MH}) = \frac{\pi(\beta, \gamma, \sigma^2, \rho_{\star} \given y)Q(\rho_{\star}, \rho)}{\pi(\beta, \gamma, \sigma^2, \rho \given y) Q(\rho, \rho_{\star})} = \frac{\pi(y \given \beta, \gamma, \sigma^2, \rho_{\star})\pi(\gamma \given \sigma^2, \rho_{\star})\pi(\rho_{\star})Q(\rho_{\star}, \rho)}{\pi(y \given \beta, \gamma, \sigma^2, \rho)\pi(\gamma \given \sigma^2, \rho)\pi(\rho)Q(\rho, \rho_{\star})}.
\]
The Jacobian of the transformation $g(x) = 1 - x$ is constant and the proposal density $Q(\rho, \rho_{\star}) \propto (1 - \rho_{\star})^{-n/2}\exp\left(-\frac{||y - X\beta - \gamma||^2}{2\sigma^2(1 - \rho_{\star})}\right)$, hence $\frac{\pi(y \given \beta, \gamma, \sigma^2, \rho_{\star})Q(\rho_{\star}, \rho)}{\pi(y \given \beta, \gamma, \sigma^2, \rho)Q(\rho, \rho_{\star})} = 1$. The remaining terms in the ratio are the prior densities. The ratio $\frac{\pi(\gamma \given \sigma^2, \rho_{\star})}{\pi(\gamma \given \sigma^2, \rho)} = \left(\frac{\rho_{\star}}{\rho}\right)^{-\frac{n}{2}}\exp\left(\frac{\gamma^{\T}V_{\phi}^{-1}\gamma}{2\sigma^2}\left(1/\rho - 1/\rho_{\star}\right)\right)$ and $\pi(\rho_{\star}) = \mbox{PC}(\rho_{\star} \given \lambda_{\rho}) \propto \lambda_{\rho}\exp(-\lambda_{\rho}d(\rho_{\star}; V_{\phi}))$. Here, $d(\rho_{\star}; V_\phi) = \sqrt{2D_{KL}(p(y; \rho_{\star})||q(y))}$ and the Kullback-Leiber divergence of $p(y; \rho_{\star}) = \mbox{N}_n(y \given 0_n, \rho_{\star} V_{\phi} + (1 - \rho{\star})I_n)$ from $q(y) = \mbox{N}_n(y \given 0_n, I_n)$ is
\[
D_{KL}(p(y; \rho_{\star})||q(y)) =-\frac{n\rho_{\star}}{2} + \frac{\rho_{\star}}{2} \sum_{i = 1}^n \frac{1}{\lambda_i} - \frac{1}{2}\sum_{i =1}^n \log\left(\rho_{\star} / \lambda_i + 1 - \rho_{\star}\right).
\]
The acceptance ratio simplifies to
\[
\exp(\alpha_{MH}) = \left(\frac{\rho_{\star}}{\rho}\right)^{-\frac{n}{2}}\exp\left(\frac{\gamma^{\T}V_{\phi}^{-1}\gamma}{2\sigma^2}\left(1/\rho - 1/\rho_{\star}\right) +\lambda_{\rho}(d(\rho; V_{\phi}) - d(\rho_{\star}; V_{\phi}))\right).
\]

\begin{algorithm}[hbt!]
\caption{Metropolis-Hastings within Gibbs sampling of posterior of $(\beta, \gamma, \sigma^2, \rho)$ in BYM2 Model with flat $\beta$ prior and PC $\rho$ prior}

\SetKwFunction{updateBeta}{updateBeta}
\SetKwFunction{updateGamma}{updateGamma}
\SetKwFunction{updateSigmaSq}{updateSigmaSq}
\SetKwFunction{MHupdateRho}{MHupdateRho}
\SetKwFunction{Fcholesky}{cholesky}
\SetKwFunction{Feigen}{eigen}
\SetKwFunction{Fbacksolve}{backsolve}
\SetKwFunction{Fforwardsolve}{forwardsolve}
\SetKwFunction{simulReduce}{simulReduce}
\SetKwProg{Fn}{function}{:}{}

Given $y, X, R = \texttt{chol}(X^{\T}X), I - H,  V_\phi^{-1}, a_{\sigma^2}, b_{\sigma^2}, \lambda_\rho, \beta^{(0)}, \gamma^{(0)}, \sigma^{2(0)}, \rho^{(0)}, c_1, \ldots c_K$:\\
Set $U, D, P, \Lambda \leftarrow \simulReduce(V_\phi^{-1}, I - H)$. \\
Precompute $ m_1 \leftarrow X^{\T}y$. \\
\For{$k = 1, \ldots, K$} {
    Set $\omega_k \leftarrow U^{\T} c_k$.
}
\For{$t = 1, \ldots, T$} {
    Set $\beta^{(t)} \leftarrow \updateBeta(\gamma^{(t - 1)}, \sigma^{2(t - 1)}, \rho^{(t - 1)})$.\\
    Set $\gamma^{(t)} \leftarrow \updateGamma(\beta^{(t)}, \sigma^{2(t - 1)}, \rho^{(t - 1)})$. \\
    Set $r^{2(t)} \leftarrow ||y - X\beta^{(t)} - \gamma^{(t)}||^2$. \\ 
    Set $\sigma^{2(t)} \leftarrow \updateSigmaSq(\rho^{(t - 1)}, r^{2(t)})$. \\
    Set $\rho^{(t)} \leftarrow \MHupdateRho(\gamma^{(t)}, \sigma^{2(t)}, \rho^{(t - 1)}, r^{2(t)})$. \\
    Set $\phi^{(t)} \leftarrow \gamma^{(t)} / (\sigma^{(t)}\sqrt{\rho^{(t)}}).$\\
    \For{$k = 1, \ldots, K$} {
        Set $\Delta^{(t)}_k = \left(\omega^{\T}_k\mbox{diag}\left(\frac{1 - \rho^{(t)}}{1 + \rho^{(t)}(d_i - 1)}\right)\omega_k\right)^{-1/2}c_k^{\T} \phi^{(t)}.$
    }
}

\Fn{\simulReduce{A, B}}{
    Solve spectral decomposition $A = P\Lambda P^{\T}$ and set $A^{-1/2} \leftarrow P\Lambda^{-1/2}P^{\T}$.\\
Solve spectral decomposition $A^{-1/2}BA^{-1/2} = CDC^{\T}$ and set $U \leftarrow A^{-1/2} C$. \\
    \KwRet $U$, $D$, $P$, $\Lambda$.
}

\label{alg:gibbs_sampling}
\end{algorithm}

Algorithm~\ref{alg:gibbs_sampling} invokes updating steps in Algorithm~\ref{alg:update_steps} to draw posterior samples of $\{\beta, \gamma, \sigma^2, \rho\}$ in a Gibbs sampling scheme with Metropolis random walk and obtains samples of $\phi = \frac{\gamma}{\sigma\sqrt{\rho}}$. 
For estimating difference boundaries, Algorithm \ref{alg:gibbs_sampling} takes in $c_1, \ldots, c_K \in \R^{n}$ corresponding to boundaries of interest. Simultaneous reduction at the start of Algorithm~\ref{alg:gibbs_sampling} increases computational efficiency and facilitates Monte Carlo estimation for $\epsilon$-difference probabilities in \eqref{eqn:ME_vk_MCest}. We compute $\Delta_k = \frac{c_k^{\T}\phi}{\sqrt{c_k^{\T} \Var(\phi \given y, \sigma^{2}, \rho) c_k}}$ for each posterior sample of $\{\phi, \rho\}$ and $k = 1, \ldots K$.

\section{Construction of Difference Probabilities on Modeled Response Level}\label{sec:model_response_diffs}

Suppose that the modeled response is $Y \given \beta, \gamma, \sigma^2, \rho \sim N(X\beta + \gamma, \sigma^2(1 - \rho))$ and the prior is $\pi(\beta, \gamma, \sigma^2) = N(\gamma\given 0, \sigma^2\rho V_\phi) \times \text{IG}(\sigma^2\given a_{\sigma^2}, b_{\sigma^2})$. We begin by assuming that $0 \leq \rho \leq 1$ and positive definite $V_\phi$ are known. Denote $\gamma_\star = (\beta^{\T}, \gamma^{\T})^{\T}$. 

Consider linear combinations of the modeled response $c_k^{\T}Y$ where $c_1, ..., c_K \in \R^{n}$. Let $\gamma_\star = (\beta, \gamma)^{\T}$, and as discussed in Section \ref{subsec:ME_model}, $\gamma_\star \given \sigma^2, \rho \sim N(M_{\star}m_{\star}, \sigma^2M_{\star}), \sigma^2 \given y, \rho \sim \text{IG}(a_n, b_n)$. Next, 
\[
Y \given y, \sigma^2, \rho \sim N_n\left(\left(I_n - WV_1^{-1}W^{\T}\right)^{-1}WV_{1}^{-1}W^{\T}y, \sigma^2(1 - \rho)\left(I_n - WV_1^{-1}W^{\T}\right)^{-1}\right)
,
\]
 where $WV^{-1}_1W^{\T} = \frac{1}{2}H + (I_n - H)(B + I_n - H)^{-1}(I_n - H)$, $B = I_n - H + \frac{1 - \rho}{\rho}V^{-1}_\phi$ and $V_\phi$ is assumed to be positive definite. For $\epsilon > 0$, define the conditional $\epsilon$-difference probabilities as
\begin{equation*}
    h_k(\epsilon; \rho)  = \Pr\left(\left.\frac{|c_k^{\T}Y| / \sqrt{\sigma^2(1 - \rho)}}{\sqrt{c_k^{\T}\left(I_n - WV_1^{-1}W^{\T}\right)^{-1}c_k}} > \epsilon \;\right|\; y, \rho\right)
\end{equation*}
for $k = 1, ..., K$. Let $0 \neq c_i, c_{j} \in \R^{n}$ and set $\alpha_i = \frac{\Ep\left(c_i^{\T}Y\given y, \sigma^2, \rho\right)}{\sqrt{\Var\left(c_i^{\T}Y/ \sigma\given y, \sigma^2, \rho\right)}} = \frac{c_i^{\T}\left(I_n - WV_1^{-1}W^{\T}\right)^{-1}WV_{1}^{-1}y}{\sqrt{(1 - \rho)c_i^{\T}\left(I_n - WV_1^{-1}W^{\T}\right)^{-1}c_i}}$. We show that the rankings of these posterior probabilities are stable by proving that given two vectors $c_{i}, c_{j} \in \R^{n}, 0 < \rho < 1$, if the corresponding conditional probabilities satisfy $h_i(\epsilon; \rho) < h_j(\epsilon; \rho)$ for some $\epsilon > 0$, then $h_i(\epsilon; \rho) < h_j(\epsilon; \rho)$ for all $\epsilon > 0$. First, define
\begin{align*}
    w_i(\epsilon; \sigma^2, \rho) &= \Pr\left(\left.\frac{|c_i^{\T}Y| / \sqrt{\sigma^2(1 - \rho)}}{\sqrt{c_i^{\T}\left(I_n - WV_1^{-1}W^{\T}\right)^{-1}c_j}} > \epsilon \;\right|\; y, \sigma^2, \rho\right) \\ &= 
    \Pr\left(-\epsilon  - \frac{\alpha_i}{\sigma} \leq Z \leq \epsilon - \frac{\alpha_i}{\sigma}\right), \; Z \sim N(0, 1) \\
    &= \Phi\left(-\epsilon + \frac{\alpha_i}{\sigma}\right) + \Phi\left(-\epsilon - \frac{\alpha_i}{\sigma}\right)
\end{align*}
where $\Phi$ denotes the standard normal CDF. Let $f: \R \rightarrow \R$ be given by $f(x)= \Phi\left(-\epsilon + \frac{x}{\sigma}\right) + \Phi\left(-\epsilon - \frac{x}{\sigma}\right)$ for any $\sigma > 0$. Then, as shown in Appendix~\ref{append:prop1_proof}, for any $x, y \in \R$, $|x| > |y|$ if and only if $f(x) > f(y)$. Therefore, for any $\sigma^2 > 0$, $w_i(\epsilon; \sigma^2, \rho) > w_j(\epsilon; \sigma^2, \rho)$ if and only if $|\alpha_i(\sigma^2)| > |\alpha_j(\sigma^2)|$. Thus, if $|\alpha_i| > |\alpha_j|$, then
\begin{equation*}
    h_i(\epsilon; \rho) = \int_{0}^\infty w_i(\epsilon; \sigma^2, \rho)\pi(\sigma^2\given y)d\sigma^2 < \int_{0}^\infty w_j(\epsilon; \sigma^2, \rho)\pi(\sigma^2\given y)d\sigma^2 = h_j(\epsilon; \rho) 
\end{equation*}
for all $\epsilon > 0$. Likewise, if $h_i(\epsilon; \rho) < h_j(\epsilon; \rho)$ for all $\epsilon$, then there exists $\epsilon > 0, \sigma^2 > 0$ such that $w_i(\epsilon; \sigma^2, \rho) > w_j(\epsilon; \sigma^2, \rho)$, which implies $|\alpha_i| > |\alpha_j|$. Since the statement $|\alpha_i| > |\alpha_j|$ does not depend on $\epsilon$, we conclude that if $h_i(\epsilon; \rho) < h_j(\epsilon; \rho)$ for some given $\epsilon > 0$, then $h_i(\epsilon; \rho) < h_j(\epsilon; \rho)$ for all $\epsilon > 0$.

Similar to the $\epsilon$-difference probabilities on the effects level, one can define the corresponding marginal posterior $\epsilon$-difference probabilities as
\begin{equation*}
    v_k(\epsilon)  = \Pr\left(\left.\frac{|c_k^{\T}Y| / \sqrt{\sigma^2(1 - \rho)}}{\sqrt{c_k^{\T}\left(I_n - WV_1^{-1}W^{\T}\right)^{-1}c_k}} > \epsilon \;\right|\; y\right).
\end{equation*}

\section{Sensitivity Analysis of Difference Threshold}\label{sec:epsilon_sensitivity}

We assess the effect of different values of the difference threshold $\epsilon$ in our proposed framework for detecting spatial disparities in a simulation experiment with a generated map using 3,078 contiguous U.S. counties. 
We use \eqref{eqn:BYM2_model} and construct $V^{-1}_\phi = c(D_W - \alpha W)$ with $W$ encoding the neighbors of U.S. counties. We generate the spatial random effects $\phi$ using $\alpha = 0.99$, scaling factor $c = 0.365$, and covariates $X = (1_n, x)$, where $x \sim \mbox{N}_n(0, I_n)$ and simulate the observed data $y$ using \eqref{eqn:BYM2_model} with $\sigma^2 = 4$, $\rho = 0.93$, and $\beta = (2, 5)^{\T}$. For the simulated data, there are $K$ = 9,119 distinct pairs of adjacent counties. We investigate spatial difference boundary detection on the set of all pairs of neighboring counties, $L = \{(i, j) : i < j, i \sim j\}$. The indicator of a true difference boundary for a neighboring pair $(i, j)$ and a given difference threshold $\epsilon$ is defined as $r_{ij}(\epsilon) = \text{I}\left(\frac{|\phi_i - \phi_j|}{\sqrt{c_{ij}^{\T}\Var(\phi \given y, \sigma^2, \rho) c_{ij}}} > \epsilon\right)$, evaluated using the true values of $\phi$ and $\rho$.  

We analyze the simulated data using the BYM2 model in \eqref{eqn:BYM2_model} and the prior $\pi(\beta, \gamma, \sigma^2, \rho) = \mbox{N}_n(\gamma \given 0, \sigma^2 \rho V_\phi) \times \text{IG}(\sigma^2 \given 0.1, 0.1) \times \mbox{PC}(\rho\given .0335, V_\phi)$, where $\lambda_{\rho} = 0.0335$ is chosen such that $\mathbb{P}(\rho \leq 0.5) \approx \frac{2}{3}$. We use Algorithm~\ref{alg:gibbs_sampling} to obtain 30,000 samples from the joint posterior of $\{\beta, \gamma, \sigma^2, \rho\}$ after discarding 10,000 burn-in samples. We then compute Monte Carlo estimates of $\{v_{ij}(\epsilon)\}_{(i, j) \in L}$ using \eqref{eqn:ME_vk_MCest} and minimize the conditional entropy loss function in \eqref{eqn:cond_ME_loss} to obtain $\epsilon_{CE} = 1.685$ as the optimal difference threshold. For this difference threshold, there are $4,841$ true difference boundaries, that is $\sum_{(i, j), i\sim j} r_{ij}(\epsilon_{CE}) = 4,841$. We compute $t^{\star} = 0.855$ using \eqref{eqn:t_cutoff} with a Bayesian FDR tolerance of $\delta = 0.05$ and report 2839 county boundaries as spatial disparities. 

\begin{figure}[ht]
    \centering
    \includegraphics[width=0.8\linewidth]{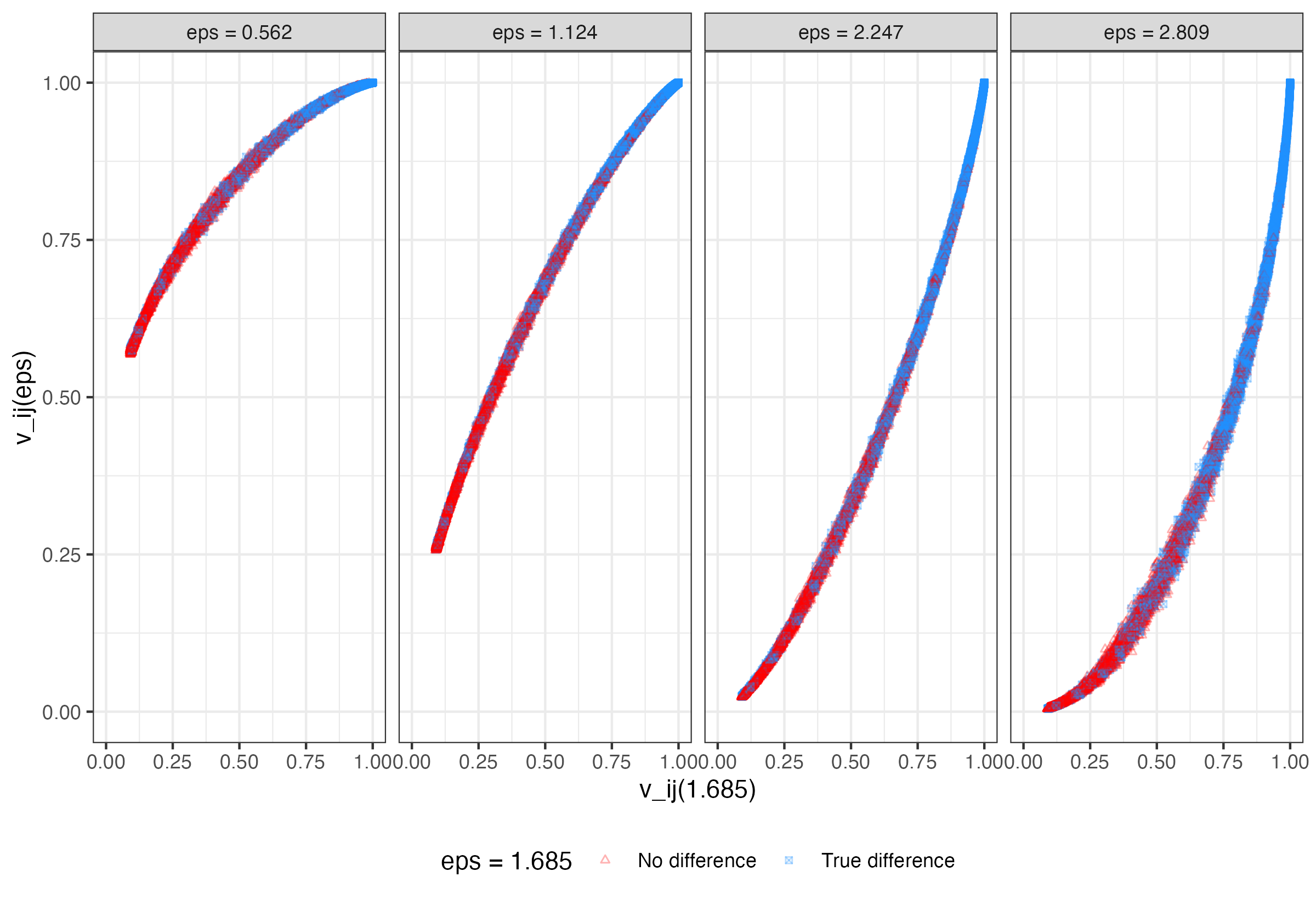}
    \caption{
    Difference probabilities $\{v_{ij}(\epsilon)\}_{(i, j) \in L}$ computed using simulated US county map and difference thresholds $\epsilon = 0.562, 1.124, 2.247, 2.809$ versus $\{v_{ij}(1.685)\}_{(i, j) \in L}$. Difference probabilities corresponding to true difference boundaries for a $\epsilon_{CE} = 1.685$ threshold are colored in blue while non-difference boundaries are colored in red.}
    \label{fig:vij_order_graph}
\end{figure}

\begin{table}[ht]
\centering
\begin{tabular}{ccccccc}
  \hline
$\epsilon$ & $t^{\star}$ & Detected Boundaries & True Boundaries & Sensitivity & Specificity & Accuracy \\ 
  \hline
0.562 & 0.811 & 6070 & 7609 & 0.742 & 0.721 & 0.739 \\ 
  1.124 & 0.832 & 4212 & 6188 & 0.638 & 0.909 & 0.725 \\ 
  $\mathbf{1.685}$ & 0.856 & 2838 & 4841 & 0.553 & 0.962 & 0.745 \\ 
  2.247 & 0.874 & 1756 & 3679 & 0.454 & 0.985 & 0.771 \\ 
  2.809 & 0.886 &  983 & 2674 & 0.352 & 0.994 & 0.806 \\ 
   \hline
\end{tabular}
\caption{Cutoff probabilities, number of detected/true difference boundaries, and classification performance with Bayesian FDR tolerance $\delta = 0.05$ across different values of difference threshold $\epsilon$. The difference threshold $\epsilon_{CE} = 1.685$ (bolded) is found from minimizing the conditional entropy loss function in \eqref{eqn:cond_ME_loss}.}
\label{tab:epsilon_sensitivity}
\end{table}

To analyze the effect of the value of $\epsilon$ on difference boundary detection, we repeat the above process for $\epsilon = 0.562, 1.124, 2.247, 2.809$. Figure~\ref{fig:vij_order_graph} shows each set of difference probabilities $\{v_{ij}(\epsilon)\}_{(i, j), i \sim j}$ plotted on the y-axis against $\{v_{ij}(\epsilon_{CE})\}_{(i, j), i \sim j}$ on the x-axis. Decreasing the difference threshold relative to $\epsilon_{CE}$ generally increases each difference probability as shown in the left two panels, while increasing the difference threshold decreases each difference probability, as seen in the right two graphs. Table~\ref{tab:epsilon_sensitivity} shows the resulting number of detected boundaries, the number of true boundaries or $\sum_{(i, j), i \sim j}r_{ij}(\epsilon)$, the sensitivity or true positive rate, the specificity or true negative rate, and the overall accuracy for each difference threshold $\epsilon$. As the difference threshold $\epsilon$ increases, the number of detected and true difference boundaries both decrease and the sensitivity decreases while the specificity increases. The entropy based choice provides a middle ground between classifying all or no boundaries as disparities.

\section{Impact of Uncertainty in Spatial Variance Proportion}\label{sec:rho_sensitivity}

We investigate the effect of uncertainty in $\rho$, the spatial proportion of variance, by reanalyzing the simulated dataset in Section~\ref{sec:epsilon_sensitivity} using a fixed value of $\rho$. We fix $\rho = 0.93$ and draw 10,000 exact samples from $\pi(\beta, \gamma, \sigma^2 \given y, \rho)$ using \eqref{eqn:aug_system_post}. For each boundary between counties, we compute a Monte Carlo estimate of the difference probability $h_{ij}(\epsilon; \rho)$ as defined in \ref{eqn:hk_in_ME_model}. Minimizing the conditional entropy loss in \eqref{eqn:cond_ME_loss} yields the optimal difference threshold of $\epsilon = 1.296$, which corresponds to 5750 true difference boundaries. For a Bayesian FDR tolerance of $\delta = 0.05$, the optimal probability cut-off point is $t^{\star} = 0.8092$, which results in reporting 3,524 pairs of neighboring counties as spatial disparities with a sensitivity of 0.583 and specificity of 0.949.

\begin{figure}[ht]
    \centering
    \includegraphics[width=0.7\linewidth]{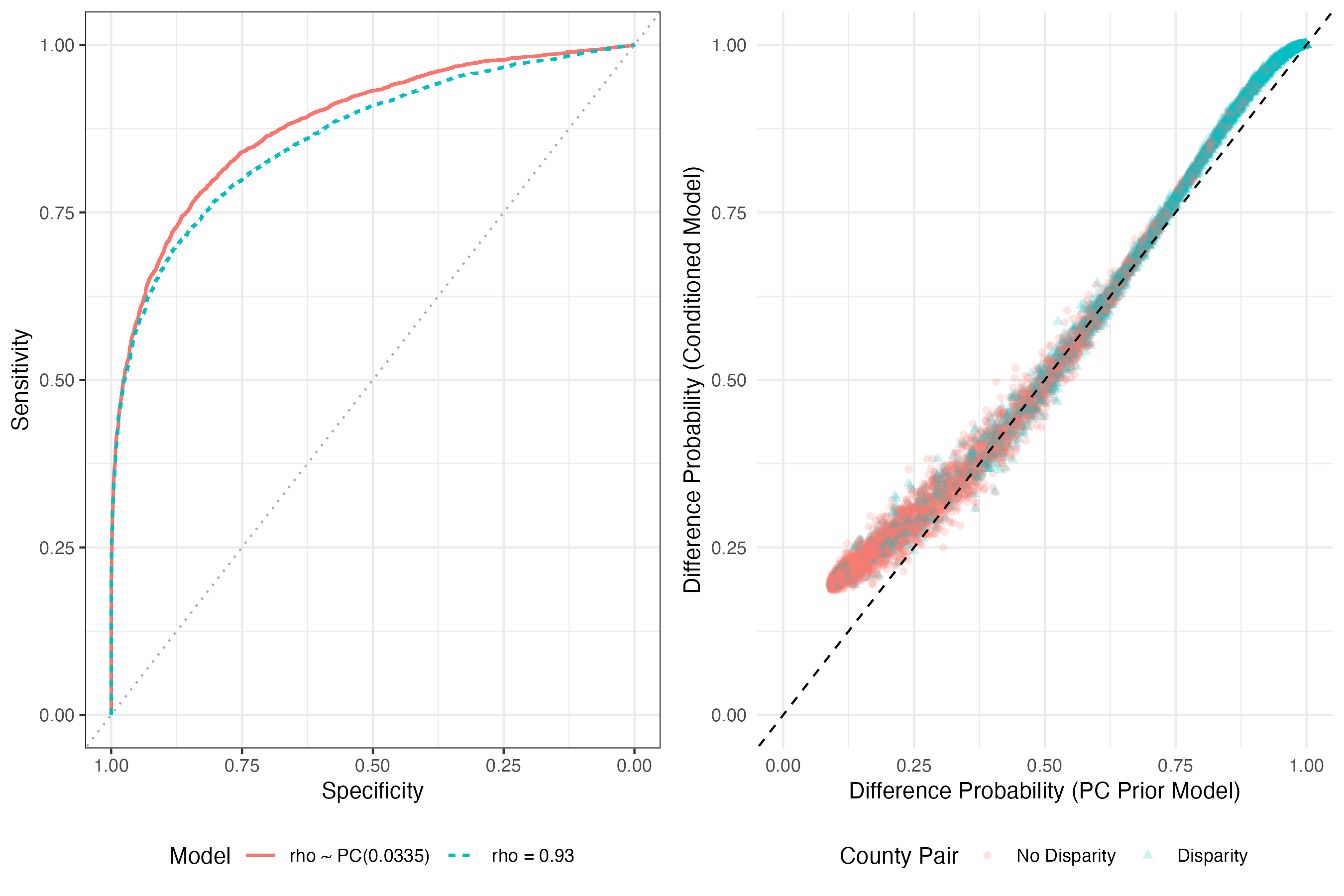}
    \caption{Left: ROC curves for simulated data analysis with a BYM2 model conditioned on $\rho = 0.93$ and difference threshold $\epsilon = 1.296$ versus analysis with a BYM2 model with a PC prior on $\rho$ and $\epsilon = 1.685$. Right: Difference probabilities conditioned on $\rho = 0.93$ versus difference probabilities from analysis with a PC prior on $\rho$. Points above the dotted line indicate neighboring pairs with difference probabilities that are greater in the conditioned model than the PC prior model.}
    \label{fig:US_sim_prob_comparison}
\end{figure}

The left panel of Figure~\ref{fig:US_sim_prob_comparison} shows the ROC curves for the model discussed in Section~\ref{sec:epsilon_sensitivity} with a PC prior on $\rho$ and difference threshold $\epsilon = 1.685$ in orange and the conditioned model with difference threshold $\epsilon = 1.296$. Although the model conditioned on the true value of $\rho$ results in reporting a greater number of difference boundaries, the model that incorporates uncertainty in the spatial proportion of variance results in slightly superior classification performance (AUC = $0.883$) than the conditioned model (AUC = $0.862$). The right panel of Figure~\ref{fig:US_sim_prob_comparison} shows a scatterplot of the difference probabilities in each model. The two sets of difference probabilities are highly correlated ($r = 0.995$) and difference probabilities corresponding to true difference boundaries are generally larger in the conditioned model. Therefore, uncertainty in $\rho$ does not appear to negatively impact classification performance but does result in more conservative classification under equivalent control of the Bayesian FDR.

\section{Classified spatial disparities in county-level tracheal, bronchus, and lung cancer mortality rates}\label{sec:RDA_disparities_table}

In Section \ref{sec:application}, we apply the maximum conditional entropy criterion to select the difference threshold $\epsilon_{CE} = 0.922$ and set the Bayesian FDR tolerance at $\delta = 0.05$ to obtain 722 out of 8,793 pairs of geographically adjacent counties within the continental US as local spatial disparities. The complete list of disparities is shown in Table~\ref{tab:RDA_disparities_table}, along with the Monte Carlo estimates of $v_k(\epsilon_{CE}) = \Pr\left(\left.\frac{|a_k^{\T}\phi|}{\sqrt{a_k^{\T}\Var(\phi \given y, \rho) a_k}} > \epsilon \;\right|\; y\right)$.

\begin{longtable}{cccc}
  \hline
 & County 1 & County 2 & $v_k(\epsilon_{CE})$ \\ 
  \hline
1 & Alachua County, Florida & Union County, Florida & 1.000 \\ 
  2 & Baker County, Florida & Union County, Florida & 1.000 \\ 
  3 & Bradford County, Florida & Union County, Florida & 1.000 \\ 
  4 & Columbia County, Florida & Union County, Florida & 1.000 \\ 
  5 & Boone County, Kentucky & Gallatin County, Kentucky & 1.000 \\ 
  6 & Carroll County, Kentucky & Gallatin County, Kentucky & 1.000 \\ 
  7 & Clark County, Kentucky & Powell County, Kentucky & 1.000 \\
\end{longtable}

\bibliographystyle{abbrvnat}
\bibliography{bayesian_spatial_disparities}